\documentclass[twocolumn,showpacs,preprintnumbers,amsmath,amssymb,superscriptaddress]{revtex4}
\usepackage{graphicx}
\usepackage{dcolumn}
\usepackage{bm}
\usepackage{longtable}

\begin{document}

\preprint{APS/123-QED}

\title{Percolation on correlated random networks}

\author{E. Agliari}
\affiliation{Dipartimento di Fisica, Universit\`a degli Studi di
Parma, Italy}
\affiliation{INFN, Sezione di
Parma}
\affiliation{Theoretische Polymerphysik, Albert-Ludwig-Universit\"{a}t, Freiburg, Germany}
\author{C. Cioli}
\affiliation{Center for Complexity Science, University of Warwick, United Kingdom}
\author{E. Guadagnini}
\affiliation{Dipartimento di Fisica, Universit\`a di
Pisa,  Italy}
\affiliation{INFN, Sezione di Pisa}
\date{\today}

\begin{abstract}
We consider a class of random, weighted networks, obtained through a redefinition of patterns in an Hopfield-like model and, by performing percolation processes, we get information about topology and resilience properties of the networks themselves. Given the weighted nature of the graphs, different kinds of bond percolation can be studied: stochastic (deleting links randomly) and deterministic (deleting links based on rank weights), each mimicking a different physical process. The evolution of the network is accordingly different, as evidenced by the behavior of the largest component size and of the distribution of cluster sizes. In particular, we can derive that weak ties are crucial in order to maintain the graph connected and that, when they are the most prone to failure, the giant component typically shrinks without abruptly breaking apart; these results have been recently evidenced in several kinds of social networks.
\end{abstract}

 \maketitle

\section{Introduction}
Network theory is a fundamental tool for the modern understanding of complex systems: by a simple graph representation, where the elementary units of a system become nodes and their mutual interactions become links, a lot of properties about the structure and dynamics of the system itself can be inferred \cite{reviews}.

Recently, the characterization of
network dynamics has become a central issue: networks are intrinsically dynamic and continuously accommodate novel members, lose their original elements, as well as build,
erase and rearrange their links \cite{libro}. The structural reorganization of networks may arise due, e.g., to a change in the resources providing the
energy to maintain their links, or to  a large stress \cite{kiss,ACV}.
In this context, percolation \cite{stauff,essam} constitutes a very interesting process able to mimic a failure or a damage of links/nodes. Moreover, percolation represents one of the simplest example of dynamical process on a graph, exhibiting a phase transitions \cite{palla} and, indeed, it has been mapped into several other critical phenomena; as well, applications in epidemiology, traffic models and in the analysis of technological networks resilience have been deeply studied \cite{appl1,appl2,appl3,appl4}.

Here we apply percolation processes as a means in order to probe the topology and the resilience of a network itself. We especially focus on 
a class of stochastic, weighted networks $\mathcal{G}$ recently introduced in \cite{AB,BA}. Such networks are  generated by assigning to each node a set of attributes and by linking two nodes whenever the pertaining attributes are similar enough; the larger the similarity, the stronger the link. As shown in \cite{AB,BA}, the resulting class of (weighted) networks $\mathcal{G}$ exhibits interesting properties such as imitative interactions (by construction), degree-degree correlation, high transitivity (i.e. a large clustering coefficient) and a properly tunable topology through a parameter $\theta$, which controls the distribution of attributes. Therefore, such networks constitute an efficient tool to describe several different systems that belong to disparate contexts, ranging from biological networks \cite{bio1,bio2}, to technological structures \cite{techno} and to social organizations \cite{socio,dissorta,fb}.

Now, since the graphs under investigation are weighted, we can perform different kinds of percolation processes: random (where links are deleted in a purely random fashion), and deterministic (where links are deleted in rank order from the weakest to strongest, or vice versa). We especially focus on graphs $\mathcal{G}$ obtained for $\theta=0$ and $\theta=0.25$, corresponding (in the limit of large size) to fully connected weighted networks and accounting for an ``unbiased'' and ``biased'' pattern distribution, respectively.
First of all, we consider the relative size of the largest connected component $S$ as a function of the fraction $f$ of links left: numerical data suggest that a ``giant component'' emerges when the fraction $f$ approaches a ``critical'' value $f_c$, which is found to scale with the system size according to $V^{-\nu}$, where $\nu$ depends on the kind of dilution. The latter also controls the sharpness of the percolation, as well as the distribution of cluster sizes, showing that, for biased pattern distributions, weak ties play a crucial role as they can be used to build up a spanning tree, conversely, strongest links are typically redundant, as a result, if weak ties are the most prone to failure the system will exhibit a poor resilience.

The paper is organized as follows: in Sec.~\ref{sec:mod} we describe the correlated random networks we are focusing on, as well as the percolation processes we perform. Then, in Sec.~\ref{sec:coup} we present the basic probability relations concerning the coupling distribution, from which we can infer qualitative information on the properties of the percolation transition; these properties are confirmed by the results of the numerical analysis that is reported in the following sections. 
In particular, the behavior of the giant component is studied in Sec.~\ref{sec:numerics}, the distribution of cluster sizes is described in Sec.~\ref{sec:cluster} and the behavior of the clustering coefficient is examined in  Sec.~\ref{sec:correlation}. An overall discussion  on the results and on the perspectives of our work is contained in Sec.~\ref{sec:conclusions}. In Appendix A we show that the class of graphs that we consider displays dissortative mixing in a wide region of the parameter set, while in Appendix B we present some analytic results which are valid for the random percolation process on $\mathcal{G}$.


\section{The model}\label{sec:mod}
We first introduce the class of networks on which we focus our analysis and later we describe the percolative processes we will perform on such networks. 

\subsection{Network generation}
Recently, a new approach to generate correlated random networks has been introduced \cite{AB,BA}; the approach is based on a simple shift $[-1, +1] \rightarrow [0, +1]$ in the definition of patterns in an Hopfield-like model and it allows to generate a broad variety of different topologies ranging from fully-connected to small-world, to extremely diluted. 

More precisely, we consider a set of $V$ nodes, each endowed with a set of $L$ attributes encoded by a binary string $\xi$; the ensemble of strings is extracted according to the probabilities $P(\xi^{\mu}=0) = (1-a)/2$ and $P(\xi^{\mu}=1) = (1+a)/2$, where the fixed parameter $a$ belongs to the interval $[-1,1]$. Then, the coupling between two generic nodes $i$ and $j$ is given by the rule 
\begin{equation}\label{eq:rule}
J_{ij} = \sum_{\mu=1}^L \xi_i^{\mu} \xi_{j}^{\mu}.
\end{equation}
Therefore, the wider the overlap between non-null entries and the larger the weight associated to the link, with $J_{ij} \in [0,L]$; the extreme case $J_{ij}=0$ means that there exists no link between nodes $i$ and $j$. The values taken by the strings  components admit the following interpretation: $\xi^\mu_i =1$ means that  agent $i$ is endowed with the particular feature $\mu$, this feature can represent a biological trait or an individual attitude according to the considered system (the absence of this particular feature corresponds to  $\xi^\mu_i =0$). Then Eq.~\ref{eq:rule} states that agents show homophily. For example, in social networks, people interact with others of similar age, income, race, etc.

As shown in \cite{AB,BA}, the way a node is connected to the network is sensitively affected by the number $\rho$ of non-null entries present in the pertaining string, that is, for the $i$-th node, $\rho_i = \sum_{\mu} \xi^{\mu}_i$ (notice that since $\rho$ is Poissonian, its average is given by $\bar{\rho} = L (1+a)/2$). In fact, one finds that the average probability
$\bar{P}_{\mathrm{link}}(\rho_i;a) $ that $i$ is connected to
another generic node, reads as
\begin{equation} \label{eq:prob2}
\nonumber
\bar{P}_{\mathrm{link}}(\rho_i;a) =  1 - \left( \frac{1-a}{2} \right)^{\rho_i}.
\end{equation}
Moreover, by averaging over all possible
string arrangements, one finds for the average link probability $p$ between two generic nodes
\begin{equation}
\nonumber
p = 1 - \left[ 1- \left( \frac{1+a}{2}\right)^2 \right]^L.
\end{equation}

The class of networks that are generated in this way exhibit different levels of correlation. For instance, it is easy to see \cite{AB,BA} that two neighbors of a given node are more likely to be connected than they would be if the graph was purely random generated; this kind of transitivity also affects the weights associated with the links \cite{AB,BA}. 
Such networks also display a dissortative behavior. Indeed, the nodes having strings with small $\rho$ typically possess a small coordination number and they are more likely to be linked with nodes with large $\rho$. The mathematical aspects of the degree correlations are elaborated in Appendix A.

Finally, we introduce the parameter $\alpha = L / V$ which turns out to crucially control not only the topology but also the thermodynamic of the system \cite{BA,AB}.
Here we assume $\alpha$ to be constant and finite, which means that, as the volume of the system grows, the length of the string increases proportionally; this corresponds the the so-called high-storage regime in neural networks \cite{amit}. 
Interestingly, as $V \rightarrow \infty$ there
exists a vanishingly small range of values for $a$ giving rise to a
non-trivial graph; such a range can be recognized by the following scaling
\begin{equation}\label{eq:scaling}
a = -1 + \frac{\gamma}{V^{\theta}},
\end{equation}
where $\theta \geq 0$ and $\gamma$ is a finite parameter.
As explained in \cite{AB}, $\theta$ controls the connectivity regime of the network- ranging from fully connected (FC, $0 \leq \theta <1/2$) to extremely diluted ($1/2 < \theta  <1$) to completely disconnected ($\theta>1$), while $\gamma$ allows a fine tuning.
In particular, here we focus on $\theta <1/2$ and $\gamma<2$, corresponding to a FC regime: in this case 
topological disorder is lost, while disorder on couplings is still present; however, notice that for $\theta = 0$ and $\gamma = 2$, the coupling distribution gets peaked at $J=L$ and disorder on couplings is relaxed as well. 

In the following, we will refer to the weighted random graph, generated as explained above, as $\mathcal{G}(\alpha,\theta,\gamma,L)$, hence highlighting the dependence on the set of parameters which control its size and its topology. We also anticipate that we will focus only on the cases $\theta=0$ and $\theta=0.25$ corresponding to weighted, complete graphs. Of course, for these cases there is no topological correlation among links (the clustering coefficient is equal to $1$ and assortativity is neutral), though correlation among link couplings is retained.

\subsection{Percolation processes}
Given an arbitrary graph, bond percolation consists in deleting the existing links with some probability $1-f$ or, in other terms, in occupying links with probability $f$; nodes connected together form clusters. When $f$ exceeds a given system-dependent threshold (or critical) value $f_c$, a macroscopic cluster, i.e. a cluster occupying a finite fraction of all available sites, also called giant component, is formed. For various network architectures and space dimensionalities this transition is typically continuous, or second-order, as the system properties changes continuously at the critical point \cite{stauff,doro}. For instance, on random networks \`a la Erd\"os-R\'enyi (ER) \cite{ER}, one starts from a set of $V$ nodes and adds links such that the probability $f$ that two nodes are joined by a link is the same for all pairs of nodes. 
When $f < 1/V$, the largest component remains
miniscule, its number of vertices scaling as $\log V$;
in contrast, if $f > 1/V$, there is a component of size
linear in $V$. Thus, the fraction of vertices
in the largest component undergoes a continuous
phase transition at $f = 1/V$.

As explained before, in the graph under study quenched weights are assigned to the edges and this allows to think of different kinds of processes, each corresponding to different physical situations: The deletion of a link may mimic the failure of the link itself due to overload \cite{ACV} or, rather, to error or attack which may affect randomly any link \cite{nature}. In the former case links with higher weight are the first to be deleted, while in the latter case deletion occurs randomly. In other kinds of situations we can think that nodes transfer a signal to neighbors and the passage of information is effective only when the tie strength is larger than some noise level \cite{immuno}. Therefore, as the level of noise grows, more and more links starting from the weakest ones, get ineffective. To summarize, we deal with the following processes:
\begin{itemize}
\item
Random percolation (RP): starting from the original graph $\mathcal{G}(\alpha,\theta,\gamma,L)$ we consider each link and we remove it with probability $1-f$, \emph{independently} of the couple of adjacent nodes, in such a way that $f$ is the fraction of links left; as $f$ is tuned from $0$ to $1$ we range from a completely disconnected graph to the original graph.
\item 
Deterministic-Weak percolation (WP): starting from $\mathcal{G}(\alpha,\theta,\gamma,L)$, we remove all links with weight \emph{smaller} than a given threshold $\iota$; that is to say, as $\iota$ is tuned from $0$ to $L$, we remove links in rank order from the weakest to strongest ties.
\item 
Deterministic-Strong percolation (SP): starting from $\mathcal{G}(\alpha,\theta,\gamma,L)$, we remove all links with weight \emph{larger} than a given threshold $\iota$; analogously to the previous case, this corresponds to remove links in rank order from the strongest to the weakest ties. 
\end{itemize}

In order to evaluate the impact of removing ties, we
measure the relative size of the largest connected component $S$, providing
the fraction of nodes that can all reach each other through
connected paths, as a function of the fraction $f$ of links left
$f$. We also measure the average squared size $\bar{S} = \sum_{s=1}^V n_s s^2/V$, 
where
$n_s$ is the number of clusters containing $s$ nodes. According to
percolation theory, if the (infinite) network collapses because of a phase
transition at $f_c$, then $\bar{S}$ diverges as $f$ approaches $f_c^{-}$ \cite{stauff,onnela}.

\section{Coupling distribution} \label{sec:coup}
The coupling distribution $P_{\mathrm{coupl}}(J; a, L)$ plays an important role as for deterministic processes, so that it is worth recalling some previous results \cite{BA} and deepening its dependence on the system parameters.

The probability for two strings $\xi_i$ and $\xi_j$ (with $\rho_i$ and $\rho_j$ non-null entries) to be connected by a link with weight $J$ is just the probability that the strings  display $J$ effective matchings; this has been found to be \cite{BA}
\begin{equation} \label{eq:fond}
P_{\mathrm{match}}(J;\rho_i,\rho_j,L) = \frac{ \binom{L}{J}  \binom{L-J}{\rho_i-J} \binom{L-\rho_i}{\rho_j-J} }{ \binom{L}{\rho_i} \binom{L}{\rho_j} },
\end{equation}
from which we can write that, in the average, the coupling distribution reads off as
\begin{eqnarray} \label{eq:coupl}
\nonumber
&&P_{\mathrm{coupl}}(J;a,L) = \sum_{\rho_i=0}^L \sum_{\rho_j=0}^L  P_{\mathrm{match}}(J;\rho_i,\rho_j,L)  \\
\times &&P_1 (\rho_i; a,L)  P_1 (\rho_j; a,L) ,
\end{eqnarray}
being $P_1 (\rho; a,L) = \binom{L}{\rho} [(1+a)/2]^{\rho} [(1-a)/2]^{L-\rho}$ the probability that a given string displays $\rho$ non-null entries.
Therefore, we get
\begin{eqnarray} \label{eq:coupl2}
P_{\mathrm{coupl}}(J;a,L) &=& \binom{L}{J} (\tilde{a}+1)^{-2L} \\
\nonumber
&\times& \sum_{\rho_i=0}^L \sum_{\rho_j=0}^L  \binom{L-J}{\rho_i -J} \binom{L - \rho_i}{\rho_2 - J} \tilde{a}^{\rho_i + \rho_j} ,
\end{eqnarray}
where we called $\tilde{a} = (1+a)/(1-a)$. Since we are focusing on the case $L = \alpha V$, with a string bias $a$ given by Eq.~\ref{eq:scaling}, it is convenient to rewrite the coupling distribution as a function of the effective parameters, namely
\begin{eqnarray} \label{eq:coupl3}
&&P_{\mathrm{coupl}}(J;\alpha,\theta,\gamma,L) = \binom{L}{J} \left[1 - \frac{\gamma}{2(L/\alpha)^{\theta}} \right]^{2L} \times\\
\nonumber
&&\sum_{\rho_i=0}^L \sum_{\rho_j=0}^L  \binom{L-J}{\rho_i -J} \binom{L - \rho_i}{\rho_j - J} \left[ \frac{\gamma}{2(L/\alpha)^{\theta} - \gamma} \right]^{\rho_i + \rho_j}.
\end{eqnarray}
The previous expression shows that for systems large enough, and $\alpha$ and $L$ fixed, the distribution gets peaked at smaller $J$ as $\theta$ is increased (when $\theta > 0.5$ only couplings with value $0$ or $1$ display non vanishing probability) and the same holds for fluctuations. Moreover, a link is absent with probability $P_{\mathrm{coupl}}(0;\alpha,\theta,\gamma,L) = [1-\gamma^2/4 (\alpha/L)^{2 \theta}]^L$, which decays to zero for $\theta<0.5$.

\begin{figure}[tb] \begin{center}
\includegraphics[width=.5\textwidth]{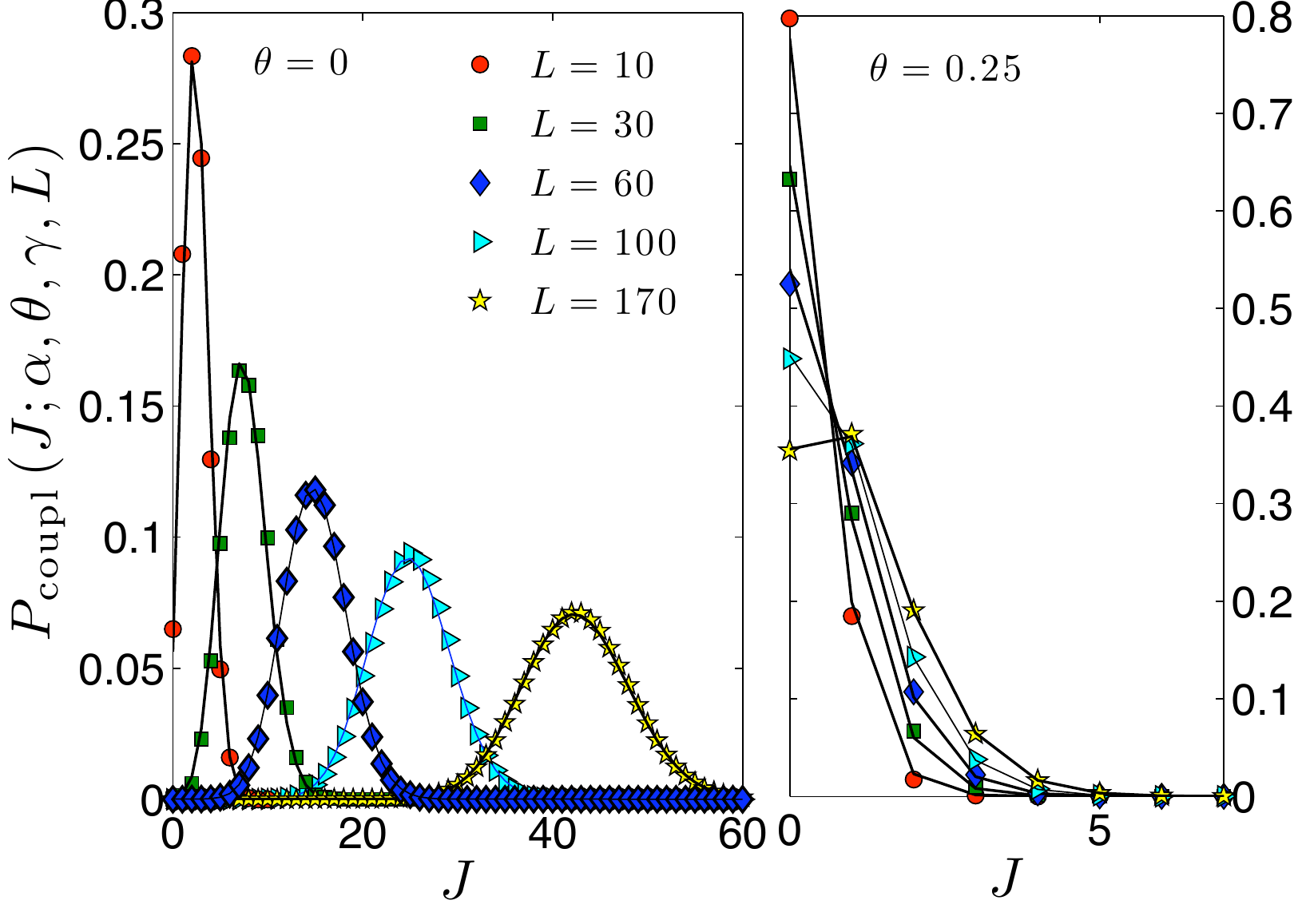}
\caption{\label{fig:distr} (Color on line) Coupling distributions $P_{\mathrm{coupl}}(J;\alpha,\theta,\gamma,L)$ from different values of $L$ (depicted in different colors, as shown by the legend) and for $\gamma=1$, $\alpha=0.1$, $\theta=0$ (left panel) or $\theta=0.25$ (right panel). Curves represent Eq.~\ref{eq:coupl3}.}
\end{center}
\end{figure}

\begin{figure}[tb] \begin{center}
\includegraphics[width=.45\textwidth]{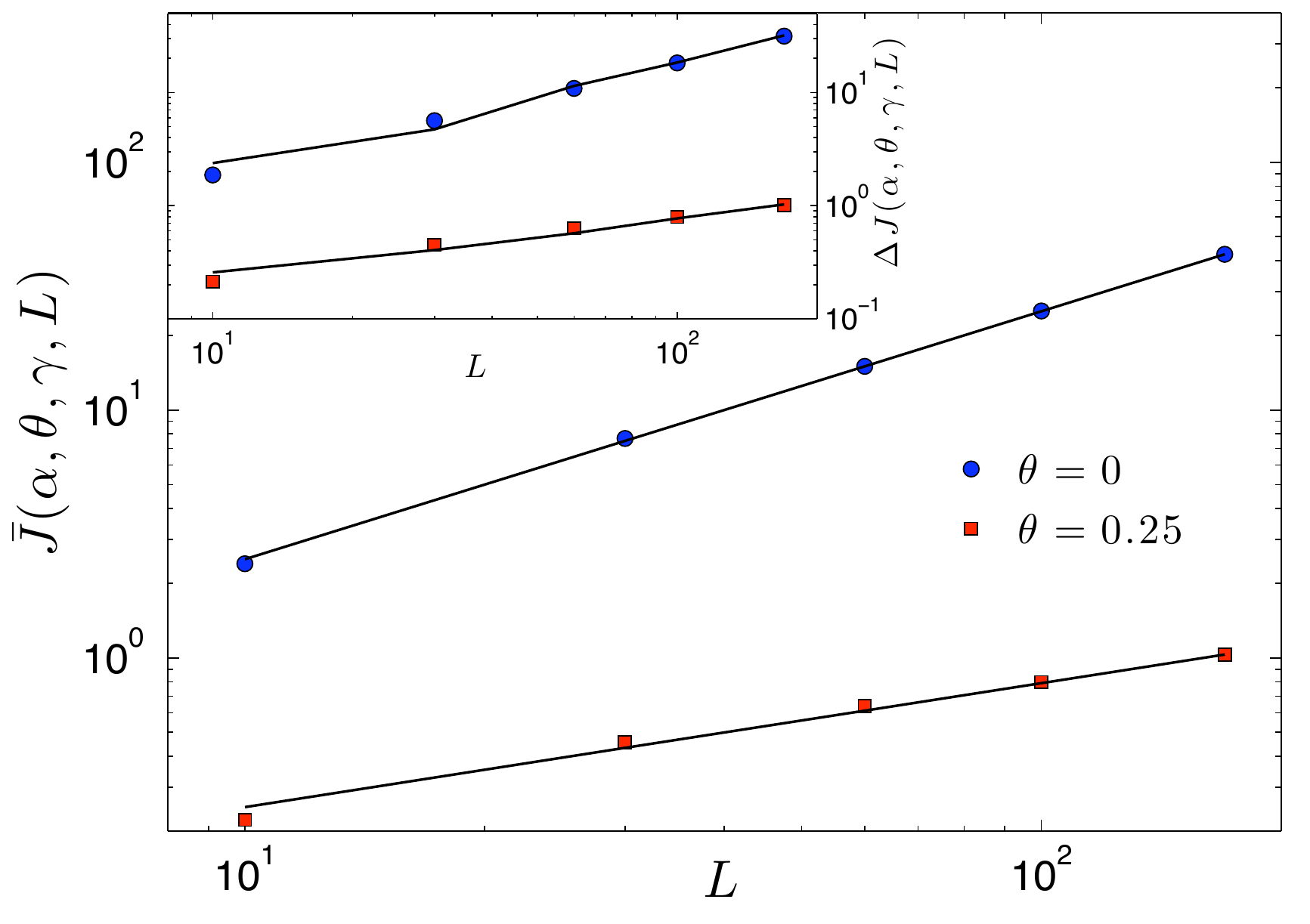}
\caption{\label{fig:mean} (Color on line) Log-log scale plot of the average coupling $\bar{J}(\alpha, \theta, \gamma, L)$ (main figure) and its fluctuations $\Delta J(\alpha, \theta, \gamma, L)$ (inset), for $\theta = 0$ ($\bigcirc$) and $\theta=0.25$ ($\square$), as shown in the legend. Symbols represent numerical data, while curves represent analytical estimates from Eq.~\ref{eq:JM} and Eq.~\ref{eq:J2M}, respectively.}
\end{center}
\end{figure}

In particular, in the following analysis we assume $\gamma =1$, $\alpha=0.1$ and $\theta =0$ or $\theta =0.25$; for $\theta=0$ we can write explicitly
\begin{eqnarray} \label{eq:coupl_0}
\nonumber
&&P_{\mathrm{coupl}}(J;0.1,0,1,L) = \frac{L!}{J!} \; 2^{-2L}  \\
&&\sum_{\rho_i=0}^L \sum_{\rho_j=0}^L  \frac{1}{(\rho_i -J)!(\rho_j-J)!(L-\rho_i-\rho_j+J)!},
\end{eqnarray}
and similarly for the latter. 
In Fig.~\ref{fig:distr} we show a comparison of the two cases where numerical data are fitted with curves given by Eq.~\ref{eq:coupl3}. Data corroborate that, even at relatively small sizes, the analytical formula above provide a good approximation and that the distribution gets broader for larger $L$ and smaller $\theta$. More precisely, we calculate the average coupling $\bar{J}(\alpha, \theta, \gamma, L)$ and its fluctuations $\Delta J ((\alpha, \theta, \gamma, L))$ as
\begin{eqnarray}
\label{eq:JM}
\bar{J}(\alpha, \theta, \gamma, L) = \sum_{J=0}^L J P_{\mathrm{coupl}}(J;\alpha,\theta,\gamma,L)  = \frac{\gamma^2}{4} \frac{L}{(L /\alpha)^{2 \theta}},\\
\label{eq:J2M}
\Delta J(\alpha, \theta, \gamma, L) = \sum_{J=0}^L (J - \bar{J})^2 P_{\mathrm{coupl}}(J;\alpha,\theta,\gamma,L),
\end{eqnarray}
where for the closed form expression in Eq.~\ref{eq:JM} we used Eq.~\ref{eq:coupl3}.
Relevant results are shown in Fig.~\ref{fig:mean}, where, again, the comparison between analytical estimates and numerical data is successful. 

Interestingly, from the width of the distribution one can infer information about the sharpness of the deterministic percolation: a broader distribution is expected to give rise to a less sharp transition. Moreover, we notice that the case $\gamma=1$ and $\theta=0$ corresponds to $a=0$, namely it corresponds to an unbiased distribution for strings, and this yields to a rather symmetric coupling distribution: as a consequence, SP and WP are expected to behave similarly. Conversely, when the coupling distribution is not symmetric, as for $\theta=0.25$, different behaviors emerge. All these points are deepened in the next section. 

Finally, we notice that for $\theta=0.25$ relatively small sizes give rise to non-fully-connected structures, that is, the coupling probability is non-null for $J=0$. As we derived from Eq.~\ref{eq:coupl3}, 
the probability that a link is absent decreases slowly with the size and such finite-size effect gets negligible only for $V \sim 10^5$. Indeed, we find that finite-size effects enhance the skewness positivity of the distribution.

\section{Percolation transitions}\label{sec:numerics}
In the following analysis we generate the graph $\mathcal{G}(\alpha,\theta,\gamma,L)$ and, while performing a dilution process (either deterministic or random), we measure the number of clusters and their size; such results are then averaged over $10^2$ realizations of $\mathcal{G}(\alpha,\theta,\gamma,L)$ in order to account for the stochasticity of the graph itself. As explained in the previous section, in the thermodynamic limit both $\theta=0$ and $\theta=0.25$ give rise to fully connected structures, so that, for large enough sizes, the random percolation process recovers the well-known results holding for ER graphs \cite{bela}.

\begin{figure}[tb] \begin{center}
\includegraphics[width=.45\textwidth]{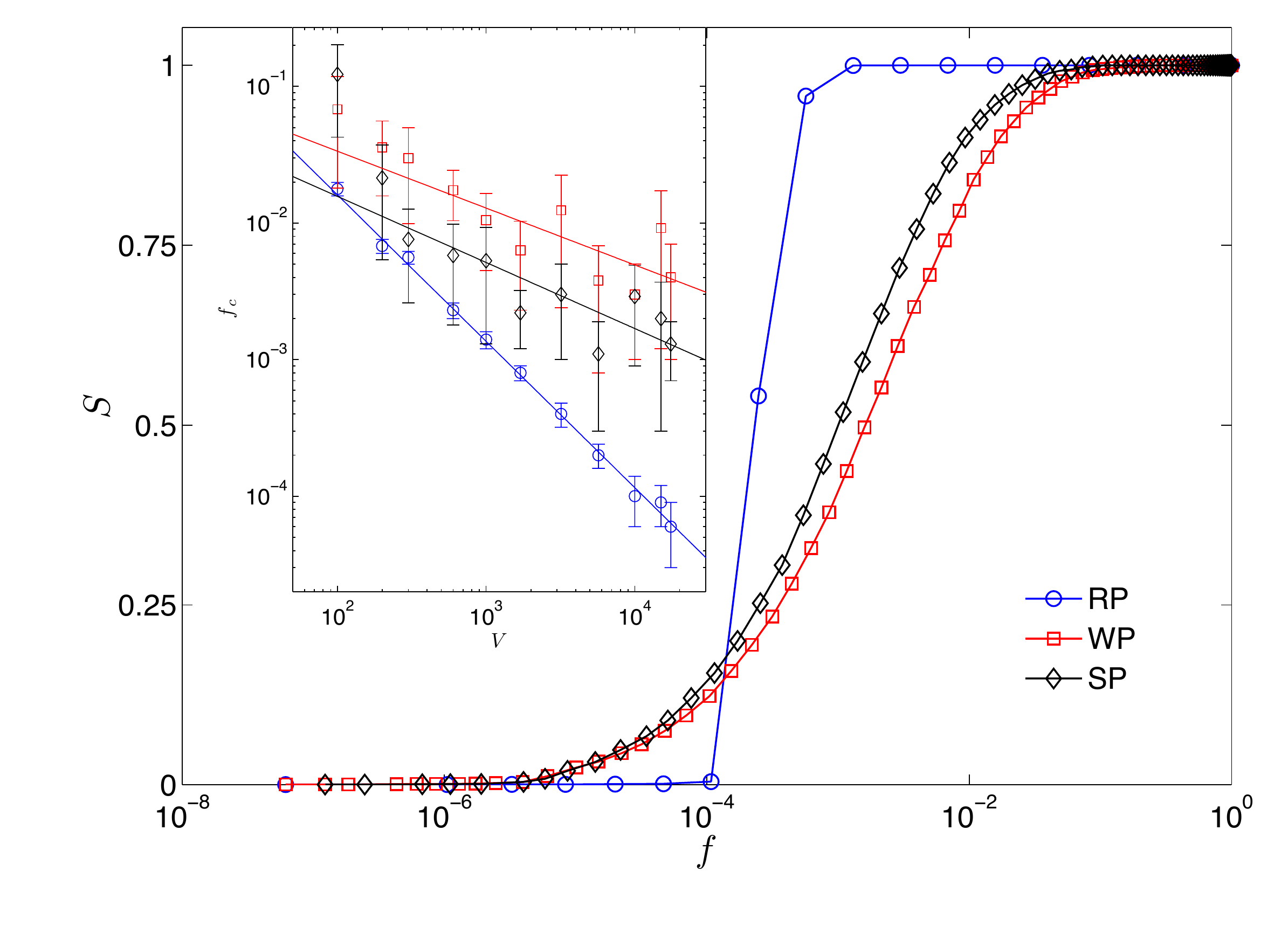}
\caption{\label{fig:Giant_0} (Color on line) Main figure: Relative size of the largest connected component $S$ versus the fraction of links left $f$ for a system of size $V = 5700$ and $\theta=0$. Fluctuations on these data, obtained by averaging over several realizations of the structure, are approximately $4 \%$. Inset: $f_c$ versus system size; symbols represent numerical data, while curves represent the best fit given by a power-law with exponents $\nu_{RP} \approx 1$ and $\nu_{WP} \approx \nu_{SP} \approx 0.5$. Different percolation processes are compared as shown by the legend.}
\end{center}
\end{figure}

\begin{figure}[tb] \begin{center}
\includegraphics[width=.45\textwidth]{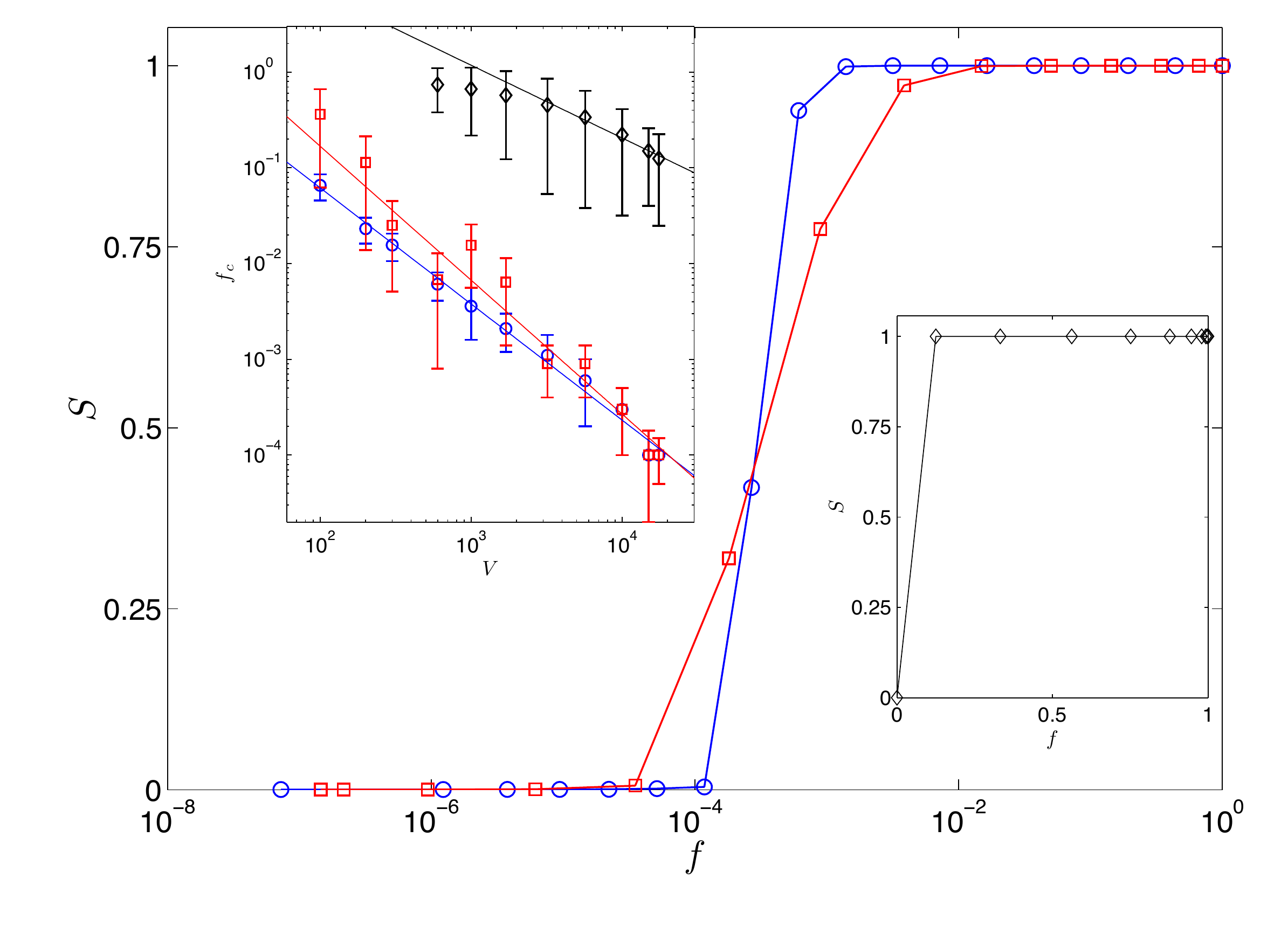}
\caption{\label{fig:Giant_025} (Color on line) Main figure: Relative size of the largest connected component $S$ versus the fraction of links left $f$ for a system of size $V = 5700$ and $\theta=0.25$. Fluctuations on these data, obtained by averaging over several realizations of the structure, are approximately $4 \%$. Different percolation processes are compared (the legend is the same as in Fig.~\ref{fig:Giant_0}). The inset on the right highlights the case of SP, where a jump in $S$ occurs around $f \sim 0.1$; notice that this jump results from the fact that $f$ cannot be continuously tuned (see also Fig.~\ref{fig:distr}). Inset on the left: $f_c$ versus system size; symbols represent numerical data, while curves represent the best fit given by a power-law with, in particular, $\nu_{RP}\approx 1$. }
\end{center}
\end{figure}

In order to evaluate the impact of removing ties, we 
measure the relative size of the largest connected component $S$ as a function of the fraction of links left $f$.
Results obtained for $\theta=0$ and $\theta=0.25$ are shown in Fig.~\ref{fig:Giant_0} and Fig.~\ref{fig:Giant_025}, respectively.

Let us comment results of Fig.~\ref{fig:Giant_0}.
First of all, we notice that when weak links are deleted first the graph starts to be disconnected ($S<1$) at a value of $f$ rather large, that is, weak ties are crucial to maintain the overall connection of the graph and, in this sense, they work as bridges. Moreover, the WP transition is smoother than the one obtained from a random deletion of edges. This suggests that the deletion of weak ties yields the disconnection, from the giant component, of single nodes (indeed those displaying small $\rho$) or of small clusters. Otherwise stated, as $f$ is increased from $0$ to $1$, we first connect nodes displaying large overlap, hence forming a strong main component, while nodes with small $\rho$ are likely to remain isolated or to form small clusters, which are successively annexed to the giant component: the process is therefore quite gentle. 
On the other hand, when links are introduced randomly, clusters grow up in a more uniform way, so that links merging disjoint components can give rise to a faster increase in the size of the giant component.
As for the SP process, when $\theta=0$ and $\gamma=1$, strings are homogeneously distributed  ($a=0$), so that, as mentioned above, no qualitative differences are expected between SP and WP; in particular, in this peculiar case ($a=0$) strong links also turn out to be crucial in maintaing the graph connected: being $\tilde{f}$ the largest fraction of links for which $S<1$, we get $\tilde{f}_{RP} < \tilde{f}_{SP} < \tilde{f}_{WP}$.

Let us now consider results for $\theta=0.25$ shown in Fig.~\ref{fig:Giant_025}. As for the RP, slight quantitative changes in $S(f)$ with respect to the previous case are due to finite size effects, while deterministic processes (DP) are also affected by the positive skewness of the coupling distribution. More precisely, strings now display only rare non-null entries so that small components (typically made up of very close or even identical strings) can arise during a WP (this explains the sharper transition); also, strong ties are rather unlikely and their deletion does not modify the connection of the giant component so that we can derive that they are redundant, that is, they typically do not participate to the spanning tree. For the sizes considered here we now have $\tilde{f}_{SP} < \tilde{f}_{RP} < \tilde{f}_{WP}$.

As anticipated, the occurrence of a percolation transition is envisaged by a singularity in the average squared size $\bar{S}$ at $f_c$, due to a network collapse as $f$ approaches $f_c$. 
We get consistent estimates for $f_c$ by evaluating the value of $f$ corresponding to a maximum in the derivative of $S$ and to the singularity in $\bar{S}$. Results for $\theta=0$ and $\theta=0.25$ are shown in the insets of Fig.~\ref{fig:Giant_0} and Fig.~\ref{fig:Giant_025}, respectively;
the sets of values for $f_c$ have also been fitted with power law functions. Of course, for the RP the expected exponent $\nu=-1$ is recovered \cite{stauff}, while for WP and SP when $\theta=0$ we find comparable exponents $\approx 0.5$; for $\theta=0.25$ finite size effects prevent to get sound estimates for $\nu$, although a different behavior of SP with respect to the other cases is apparent.
In any case, the RP corresponds to smaller values of $f_c$, meaning that a smaller fraction of links is necessary to carry a giant component, that is, a smaller degree of redundancy is retained.

Finally, some analytical insights for the case of arbitrary $\theta$ and random dilution are presented in the Appendix B, where we show consistency with known results about non-correlated networks.

\section{Cluster distributions}\label{sec:cluster}
While previous results offer a global description of the network dynamics, in this section we focus on the evolution of the internal organization of clusters by measuring the distribution $N(\rho,s)$, representing the number of nodes corresponding to a string with $\rho$ non-null entries and belonging to a cluster of size $s$.

\begin{figure*}[tb] \begin{center}
\includegraphics[width=.21\textwidth, angle=270]{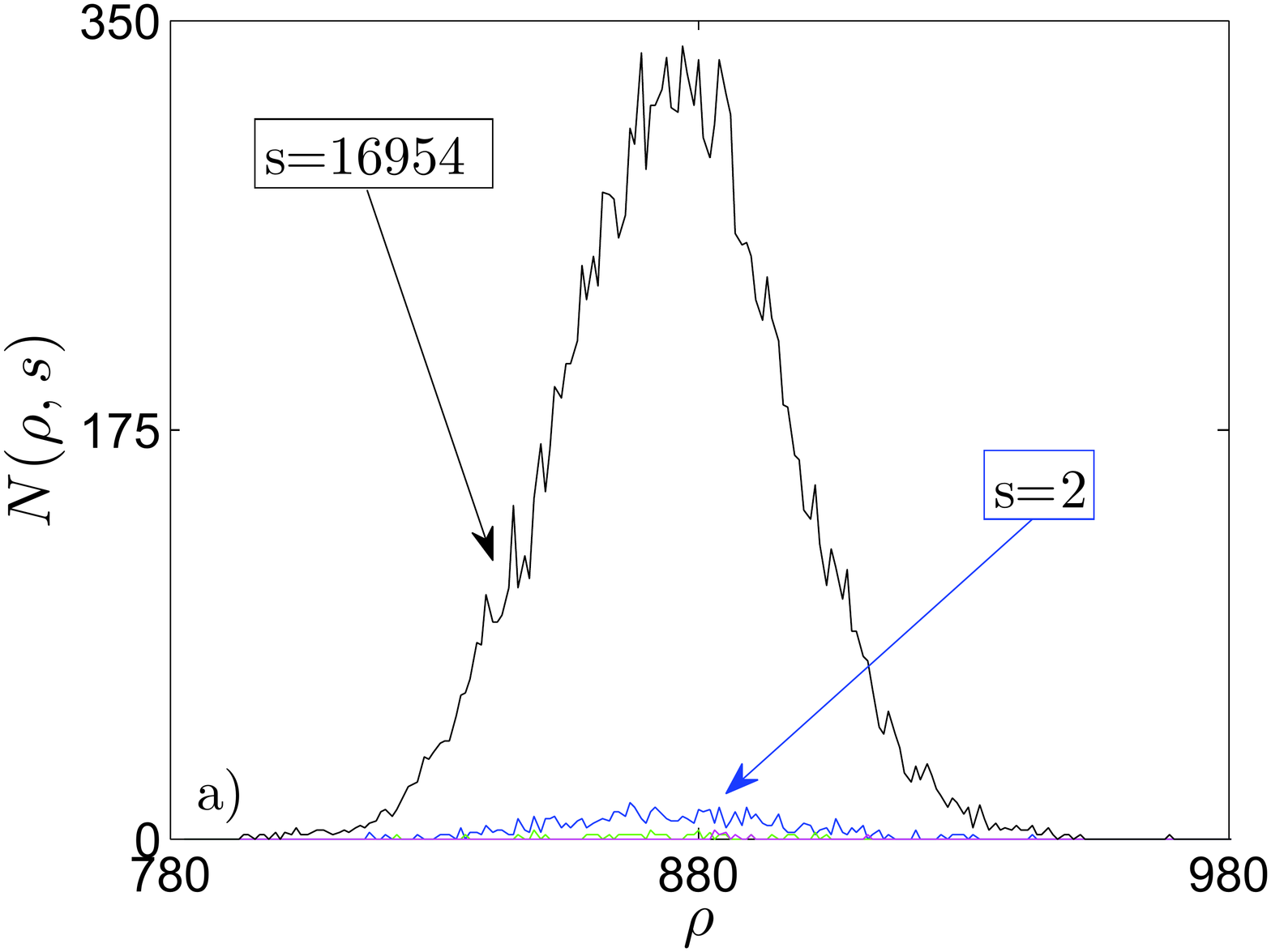}
\includegraphics[width=.21\textwidth, angle=270]{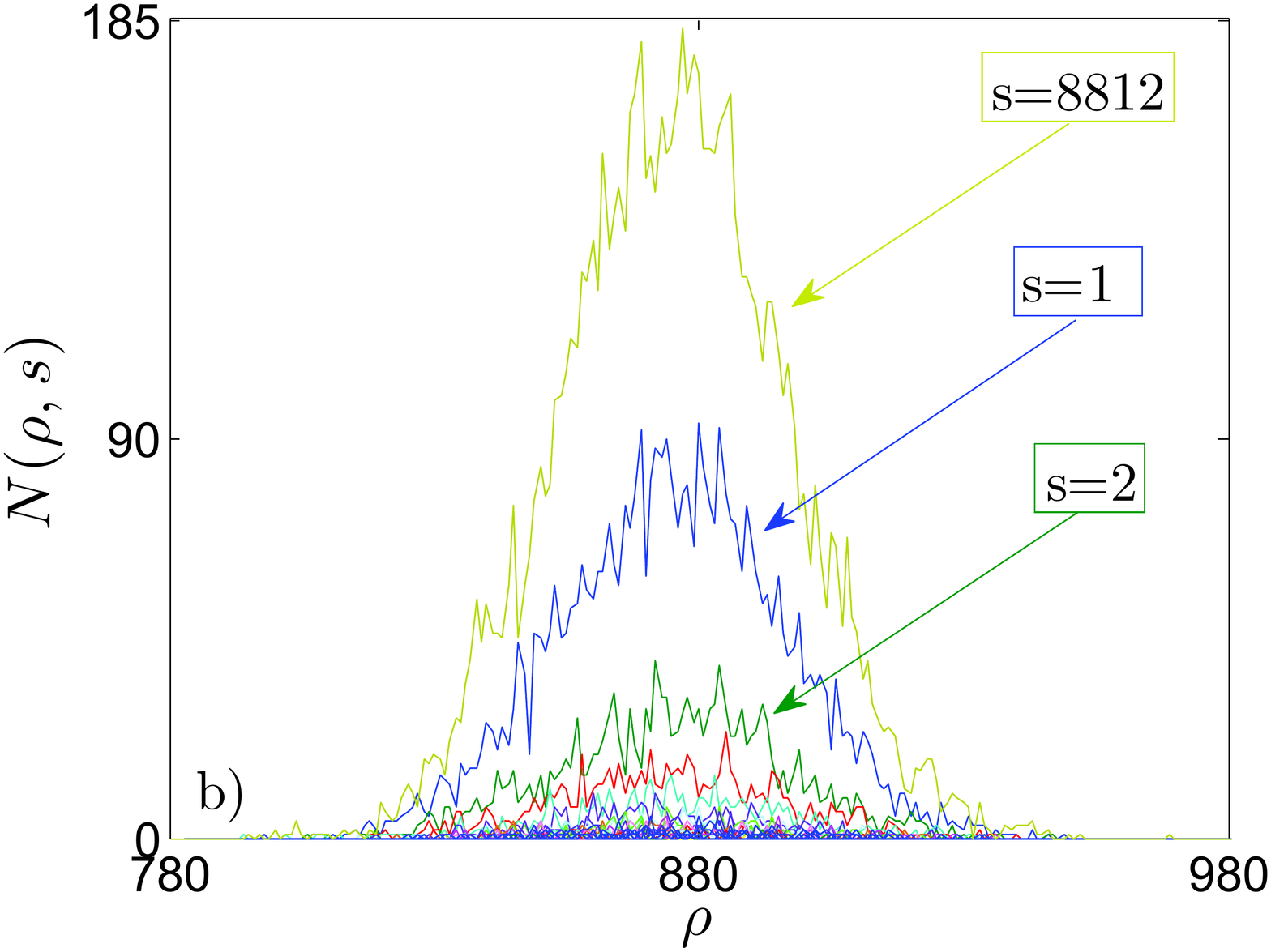}
\includegraphics[width=.21\textwidth, angle=270]{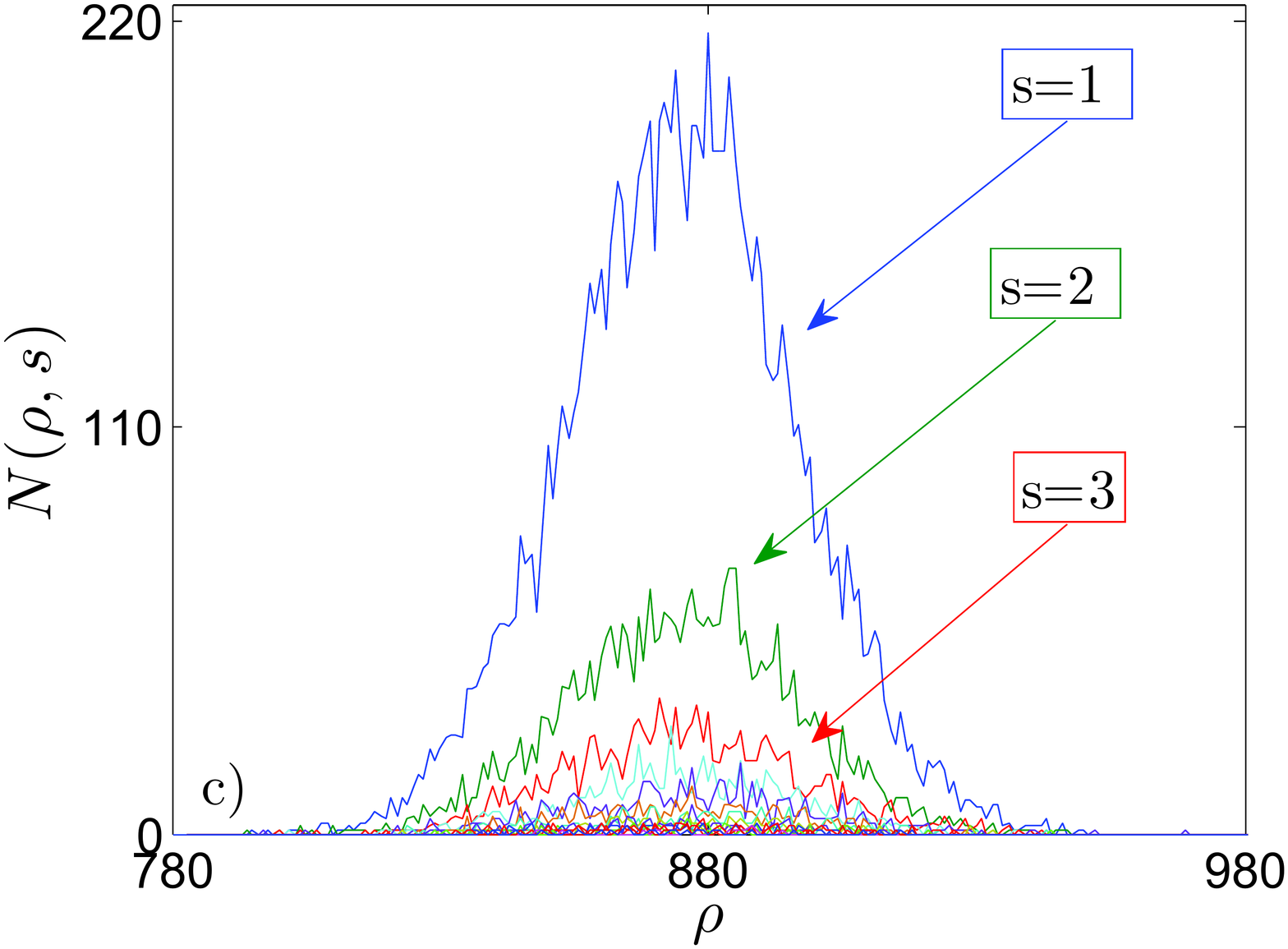}
\includegraphics[width=.21\textwidth, angle=270]{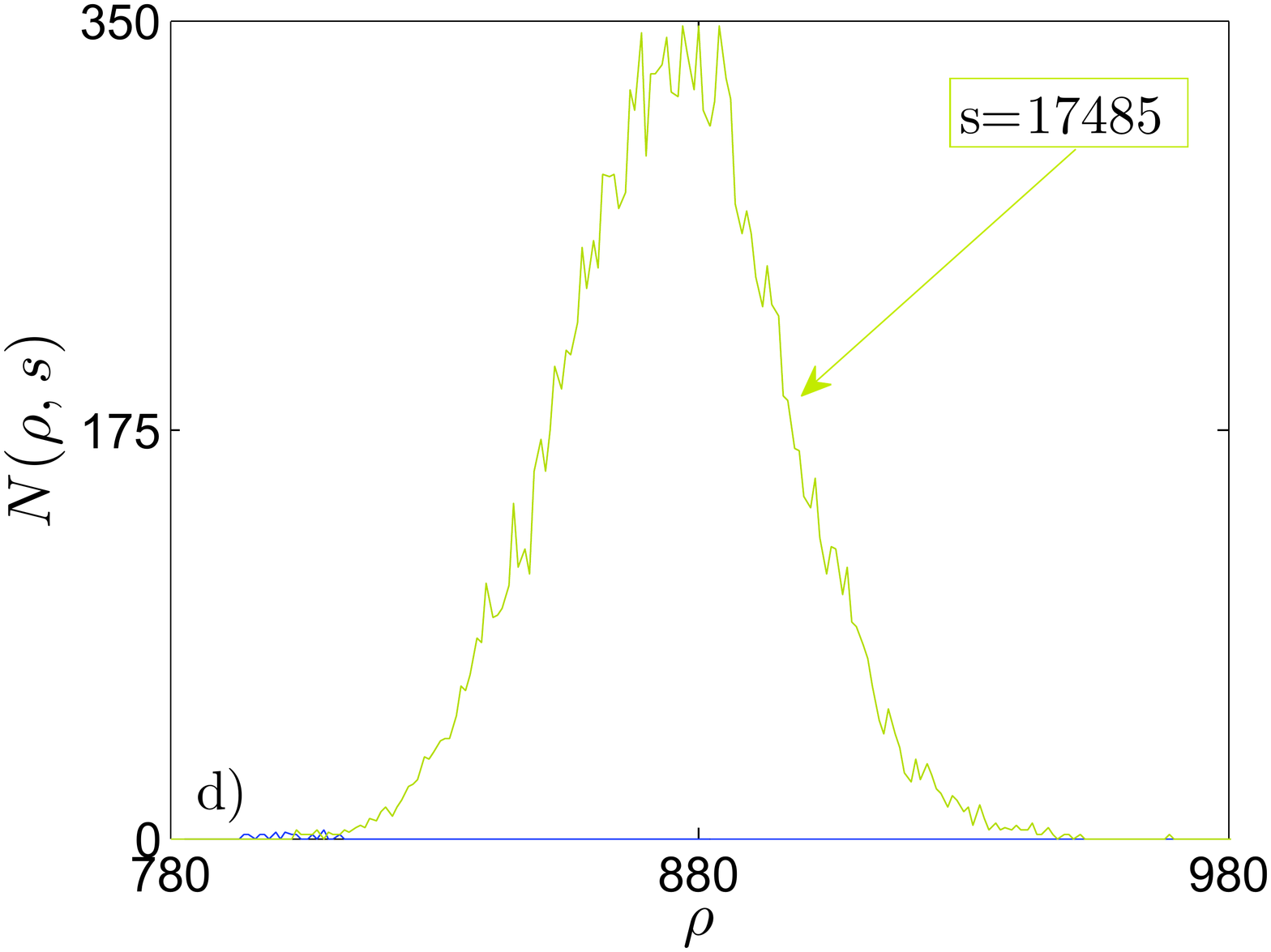}
\includegraphics[width=.21\textwidth, angle=270]{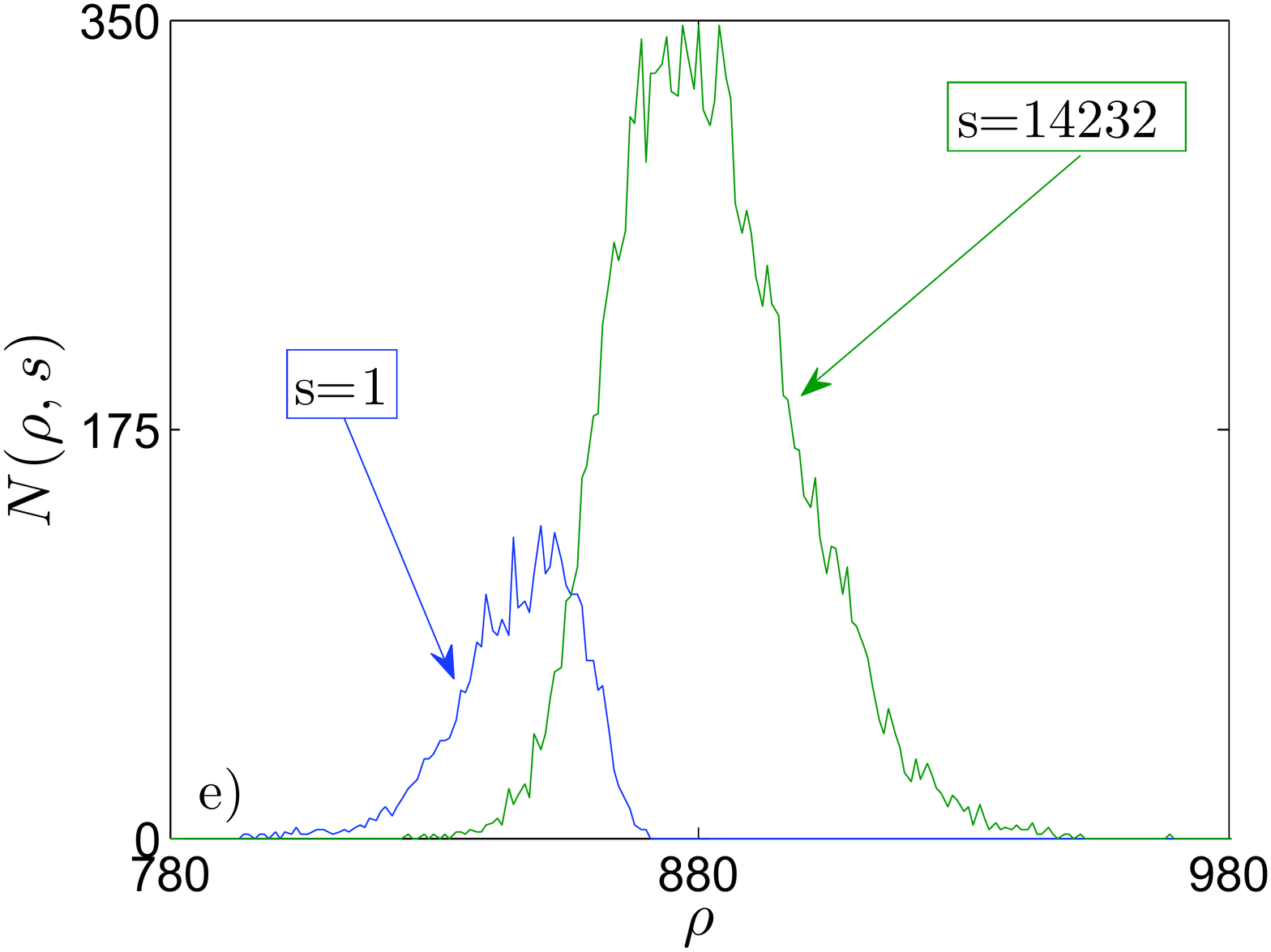}
\includegraphics[width=.21\textwidth, angle=270]{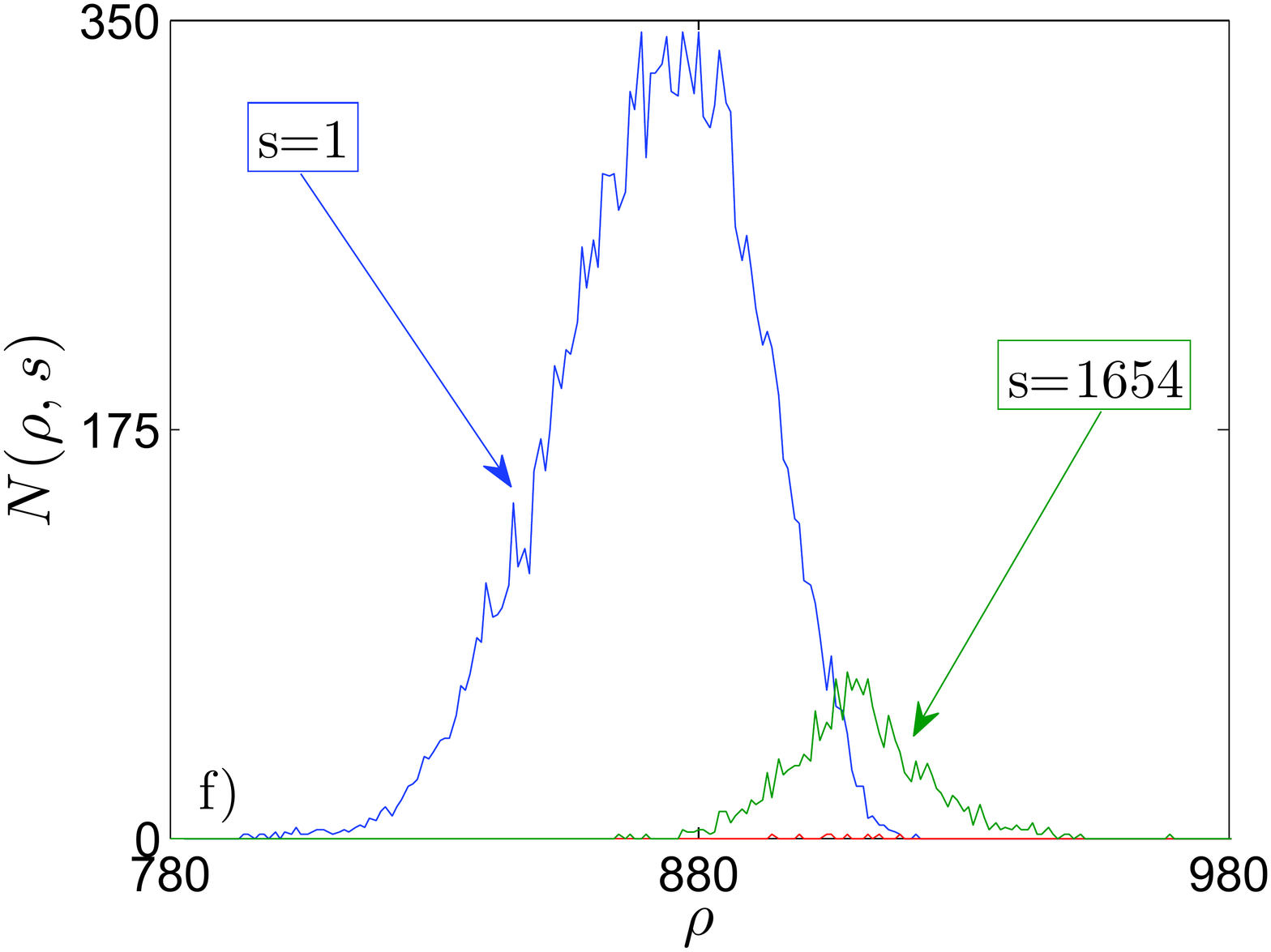}
\includegraphics[width=.21\textwidth, angle=270]{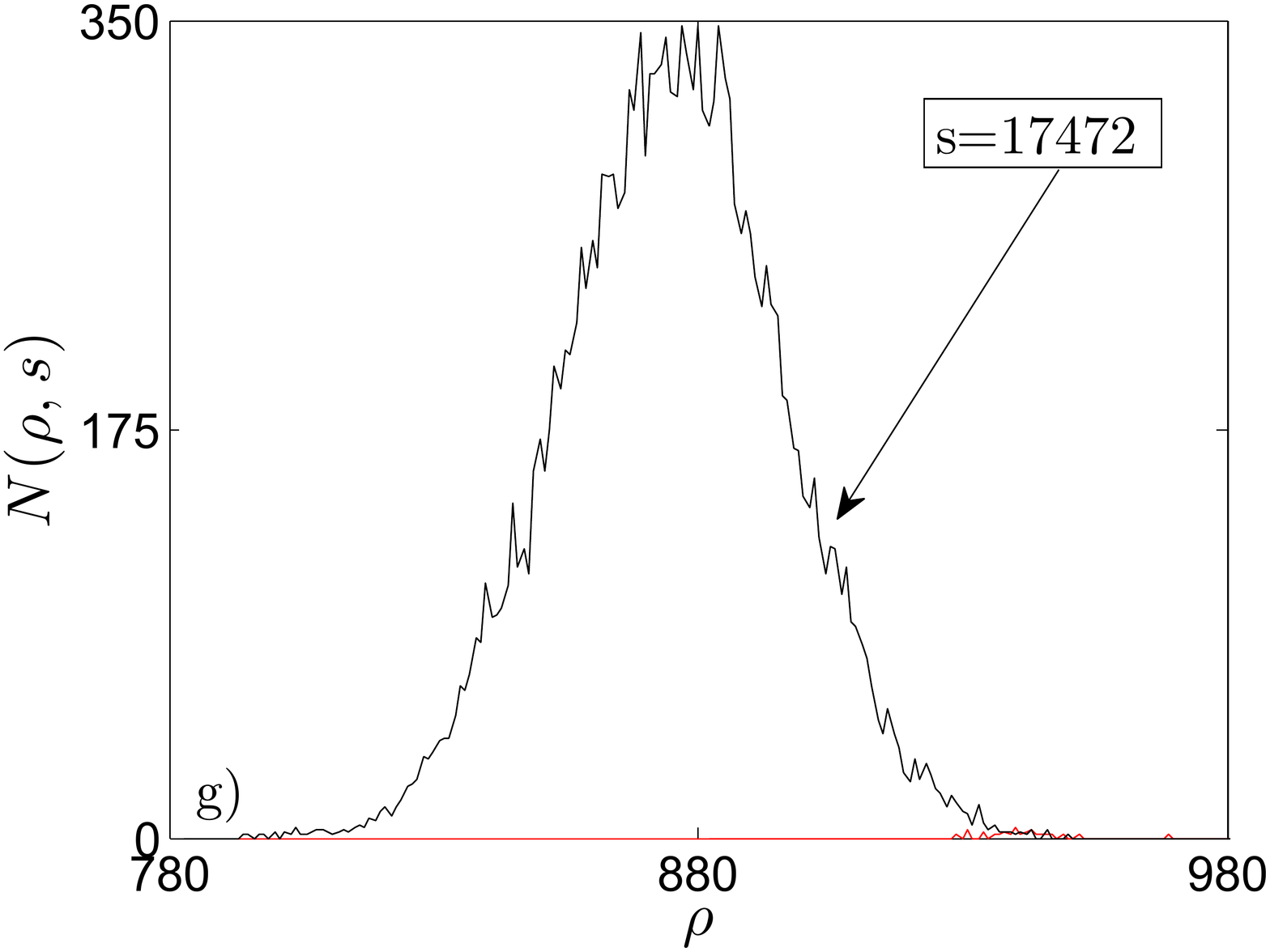}
\includegraphics[width=.21\textwidth, angle=270]{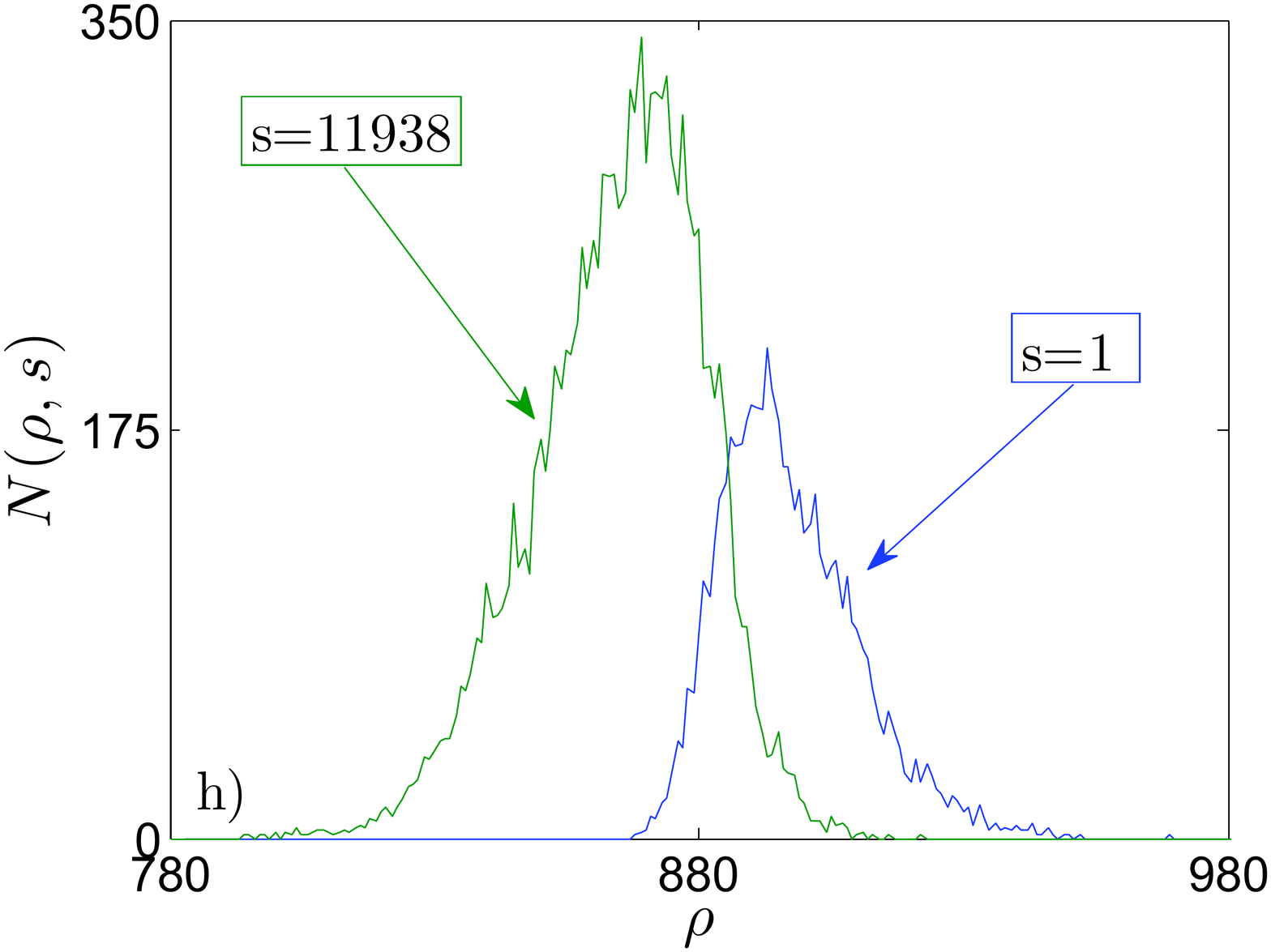}
\includegraphics[width=.21\textwidth, angle=270]{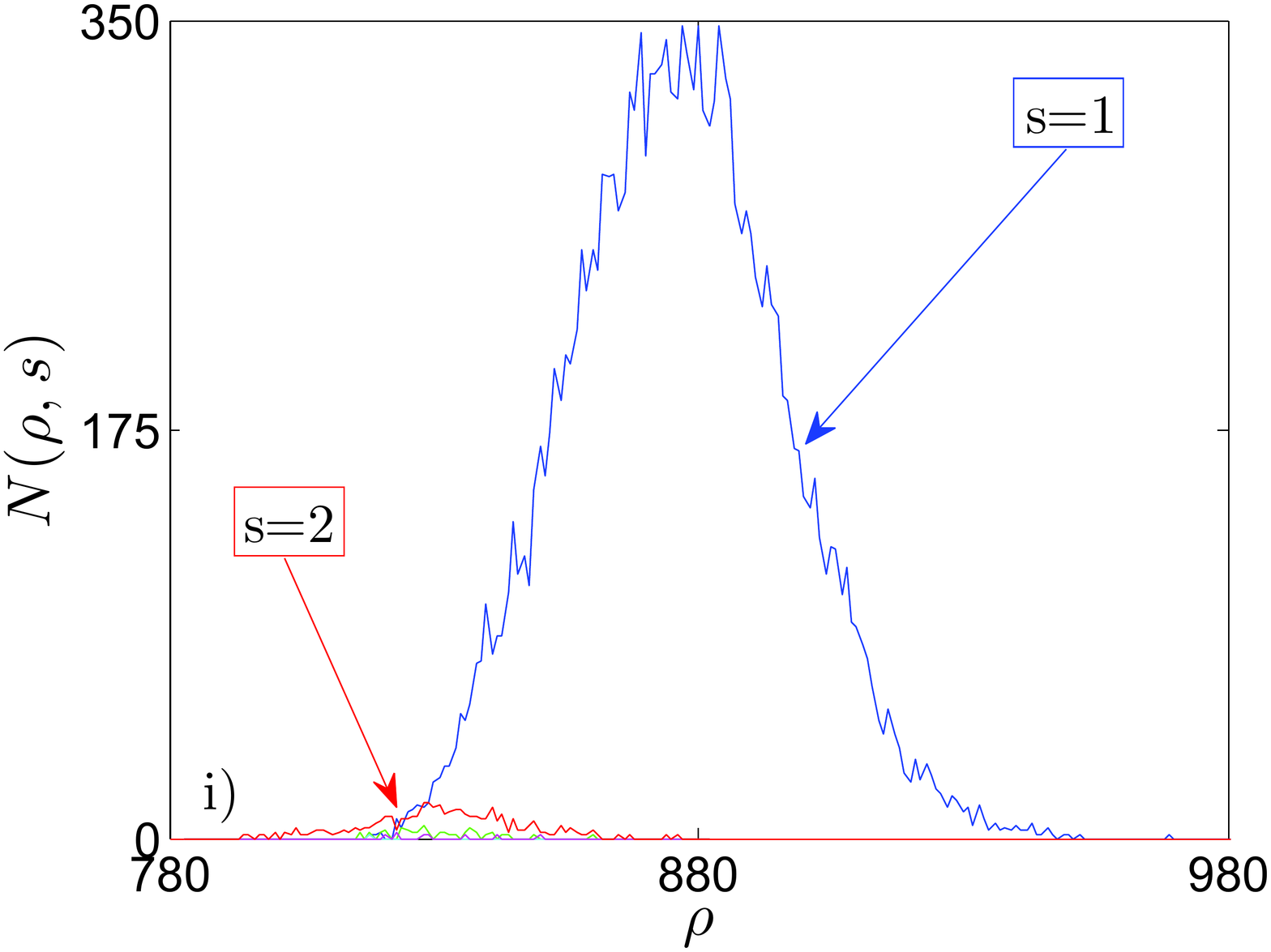}
\caption{\label{fig:Distr} (Color on line) Number of nodes $N(\rho,s)$ corresponding to a string with $\rho$ non-null entries and belonging to a cluster of size $s$, plotted as a function of $\rho$, while different values of $s$ and shown in different colors. The system considered has size $V=17500$, $\theta=0$, $\alpha=0.1$ and $\gamma=1$. The nine panels are arranged in such a way that each row represents a different dilution process (from top to bottom: RP, WP, SP), while each column represents a different regime (from left to right: $f<f_c$, $f \approx f_c$, $f>f_c$).}
\end{center}
\end{figure*}

In Fig.~\ref{fig:Distr} we show three sets of snapshots of $N(\rho,s)$ for the case $\theta=0$; each row represents a different dilution process (from top to bottom: RP, WP, SP), while each column represents a different regime (from left to right: $f<f_c$, $f \approx f_c$, $f>f_c$).
While for the random dilution intermediate and high dilution regimes (panels $b$ and $c$) are characterized by the existence of several clusters of different sizes, for deterministic dilution (panel $e$, $f$ and $h$, $i$) a node basically either belongs to the largest component or is isolated. Moreover, in the former case due to the homogeneity underlying the process, curves are all peaked at around $\bar{\rho}$ namely, the set of attributes characterizing a given node does not affect the cluster size the node belongs to. Conversely, for WP (SP), larger (smaller) values of $s$ yields distributions $N(\rho,s)$ peaked at larger values for $\rho$.
In particular, when weak ties are the most prone to failure, nodes displaying strings with large $\rho$ are the most likely to belong to the giant component.

\begin{figure*}[tb] \begin{center}
\includegraphics[width=.21\textwidth, angle=270]{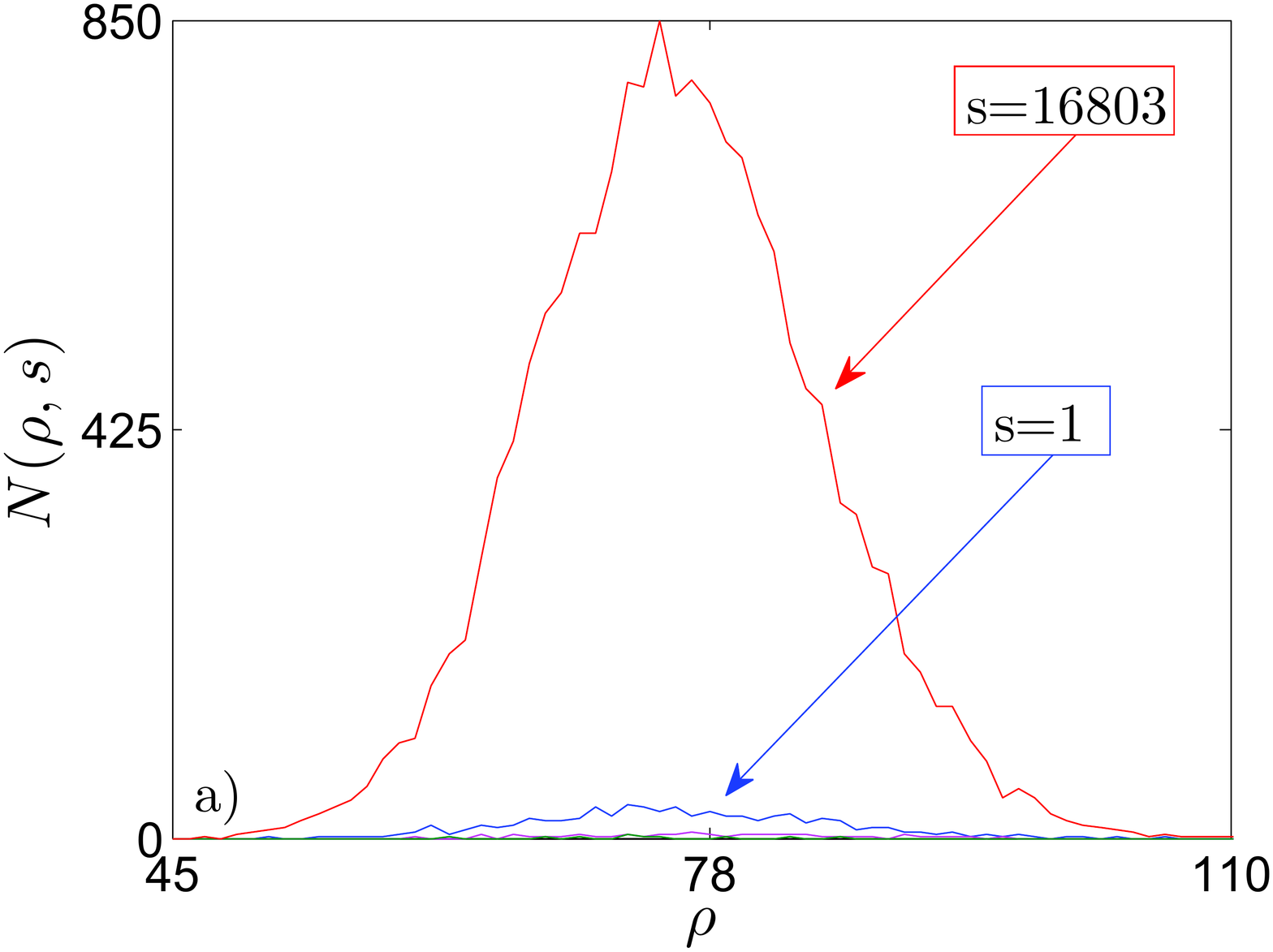}
\includegraphics[width=.21\textwidth, angle=270]{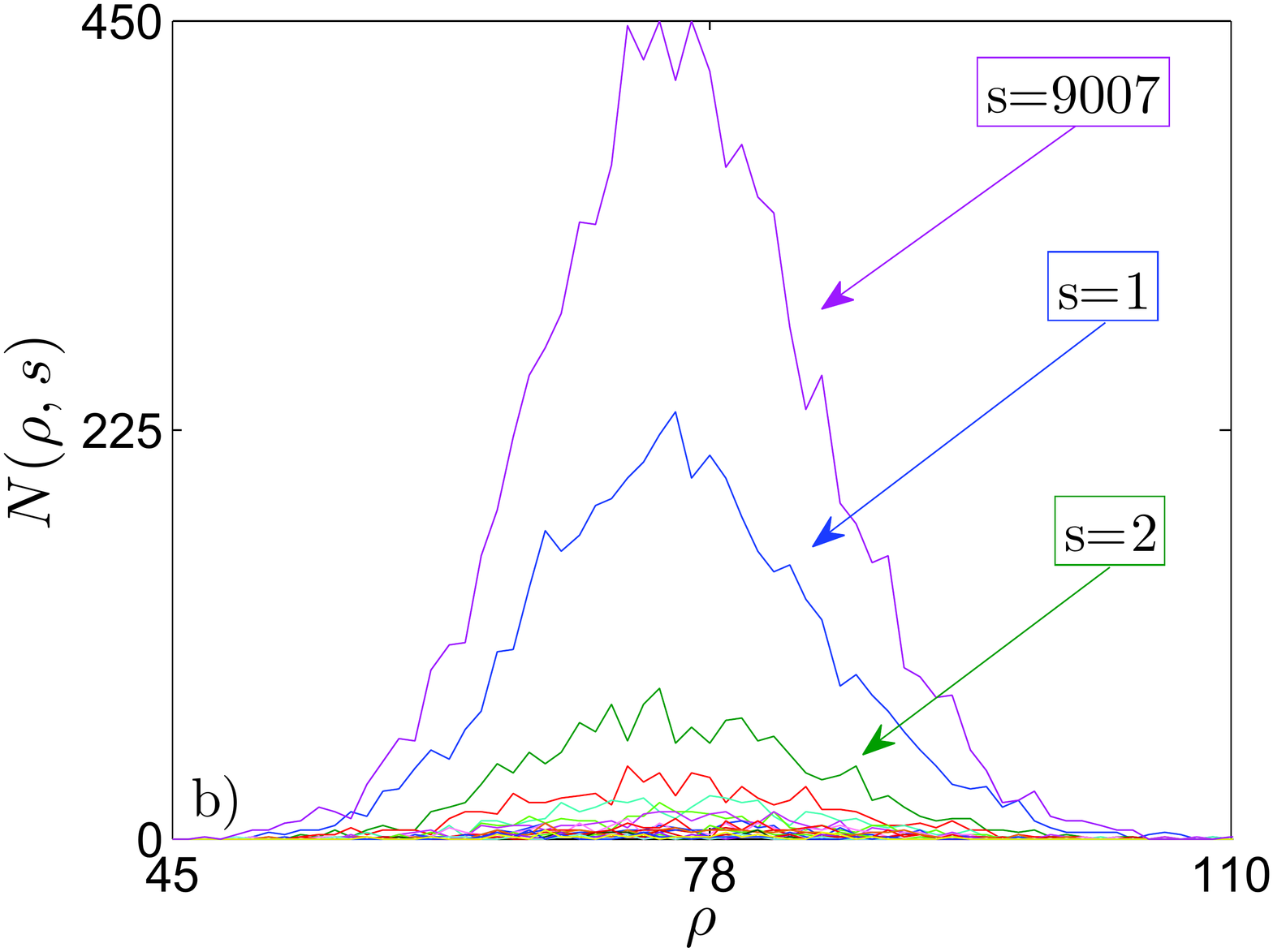}
\includegraphics[width=.21\textwidth, angle=270]{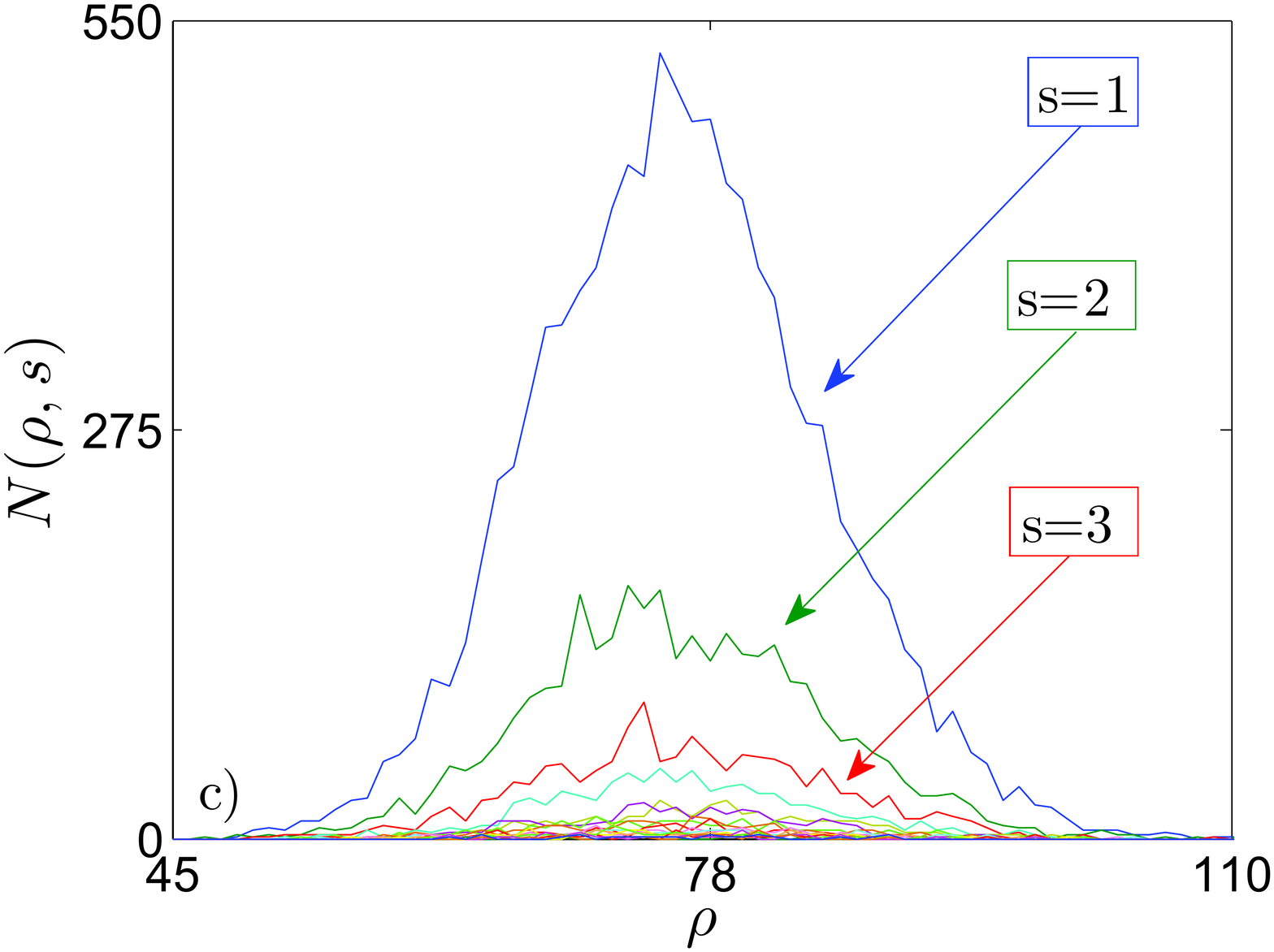}
\includegraphics[width=.21\textwidth, angle=270]{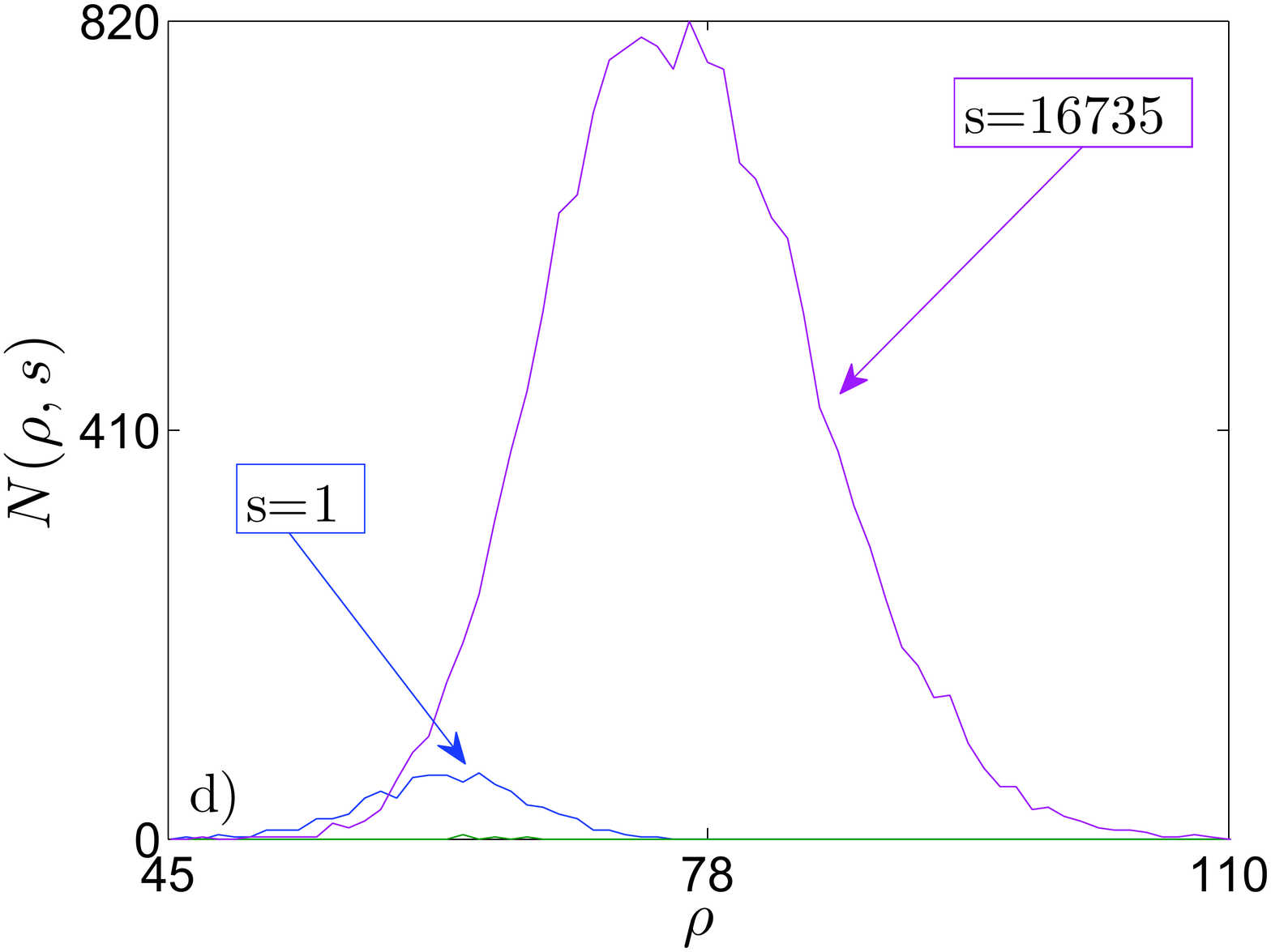}
\includegraphics[width=.21\textwidth, angle=270]{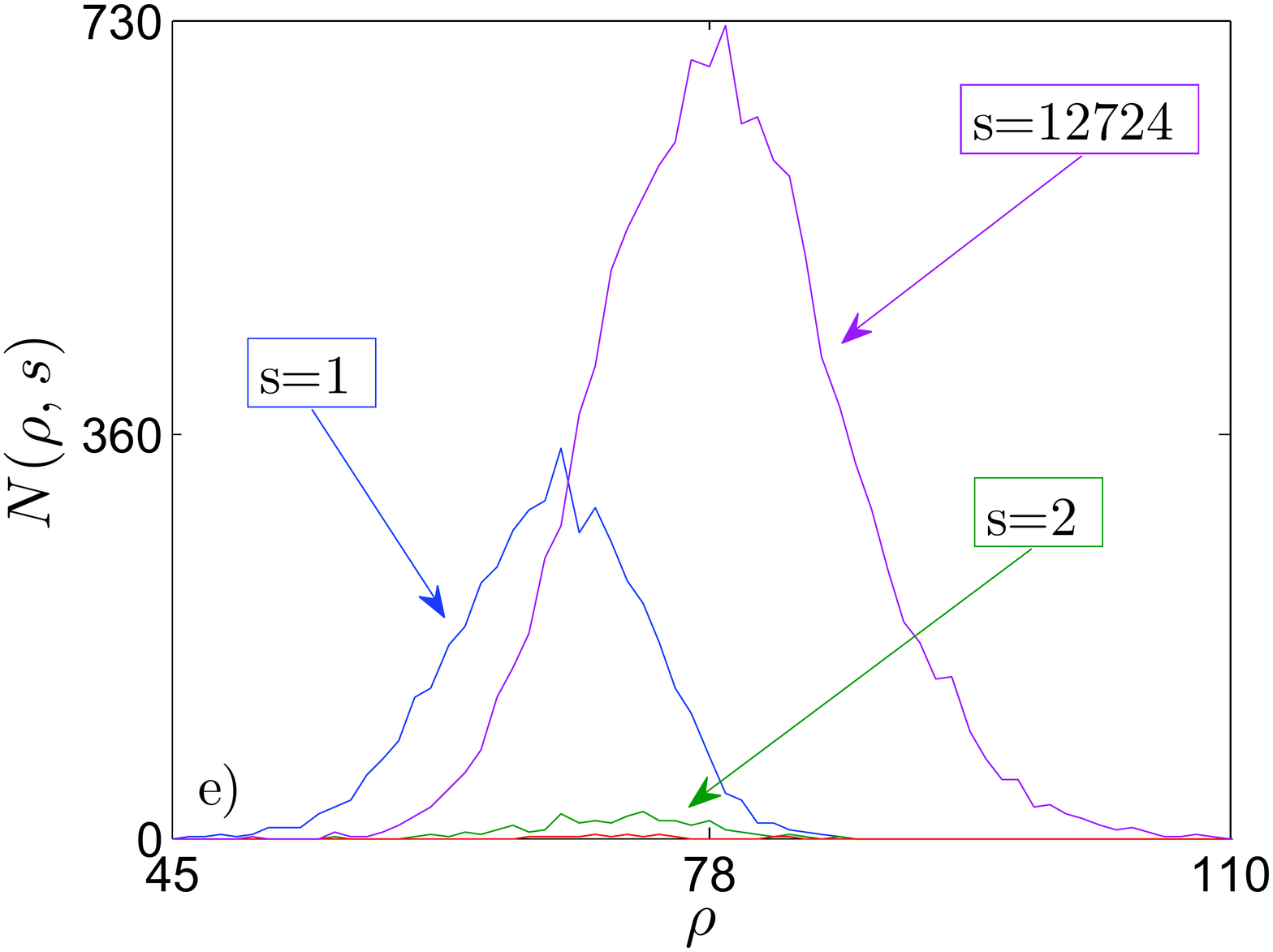}
\includegraphics[width=.21\textwidth, angle=270]{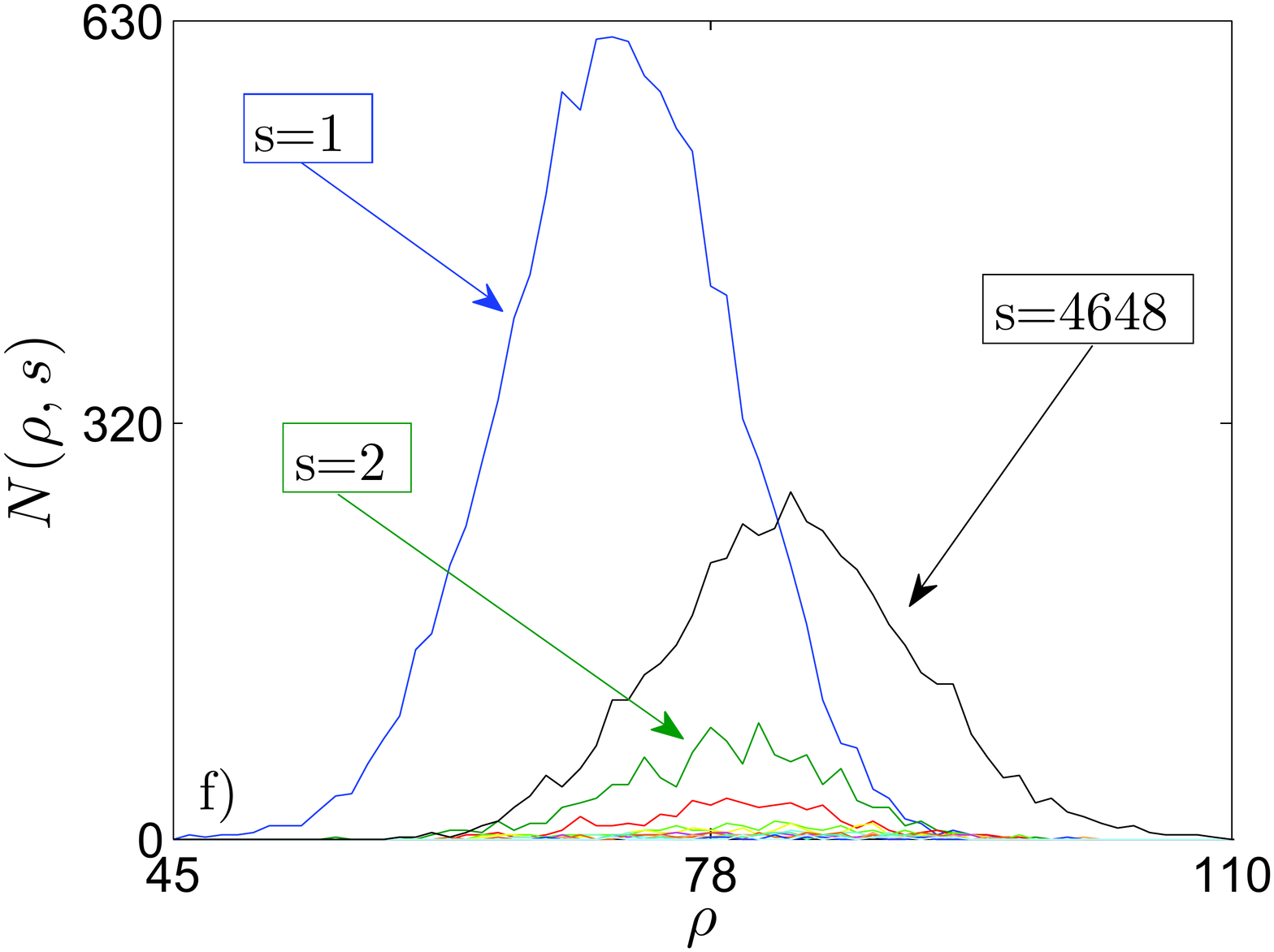}
\includegraphics[width=.21\textwidth, angle=270]{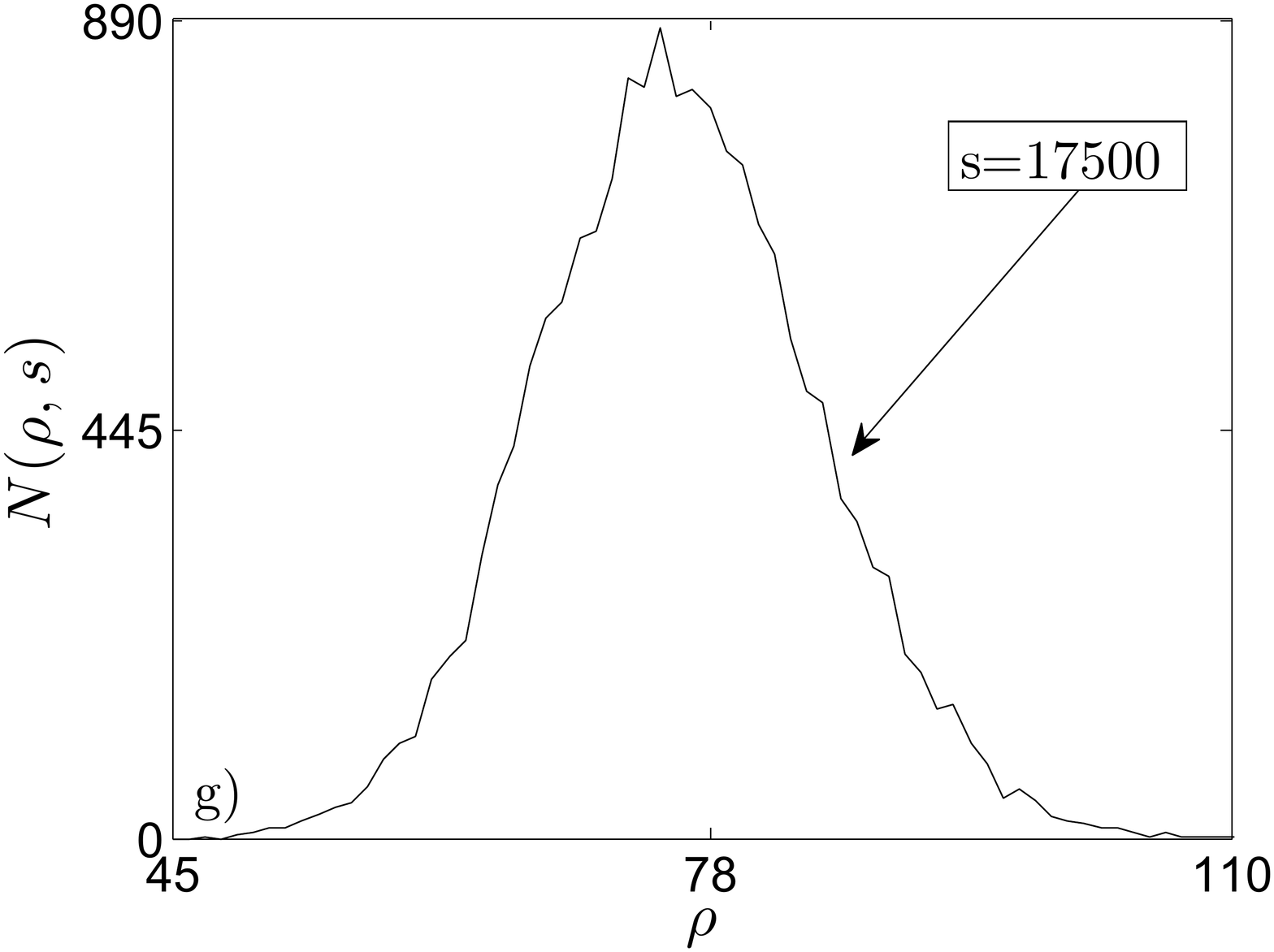}
\includegraphics[width=.21\textwidth, angle=270]{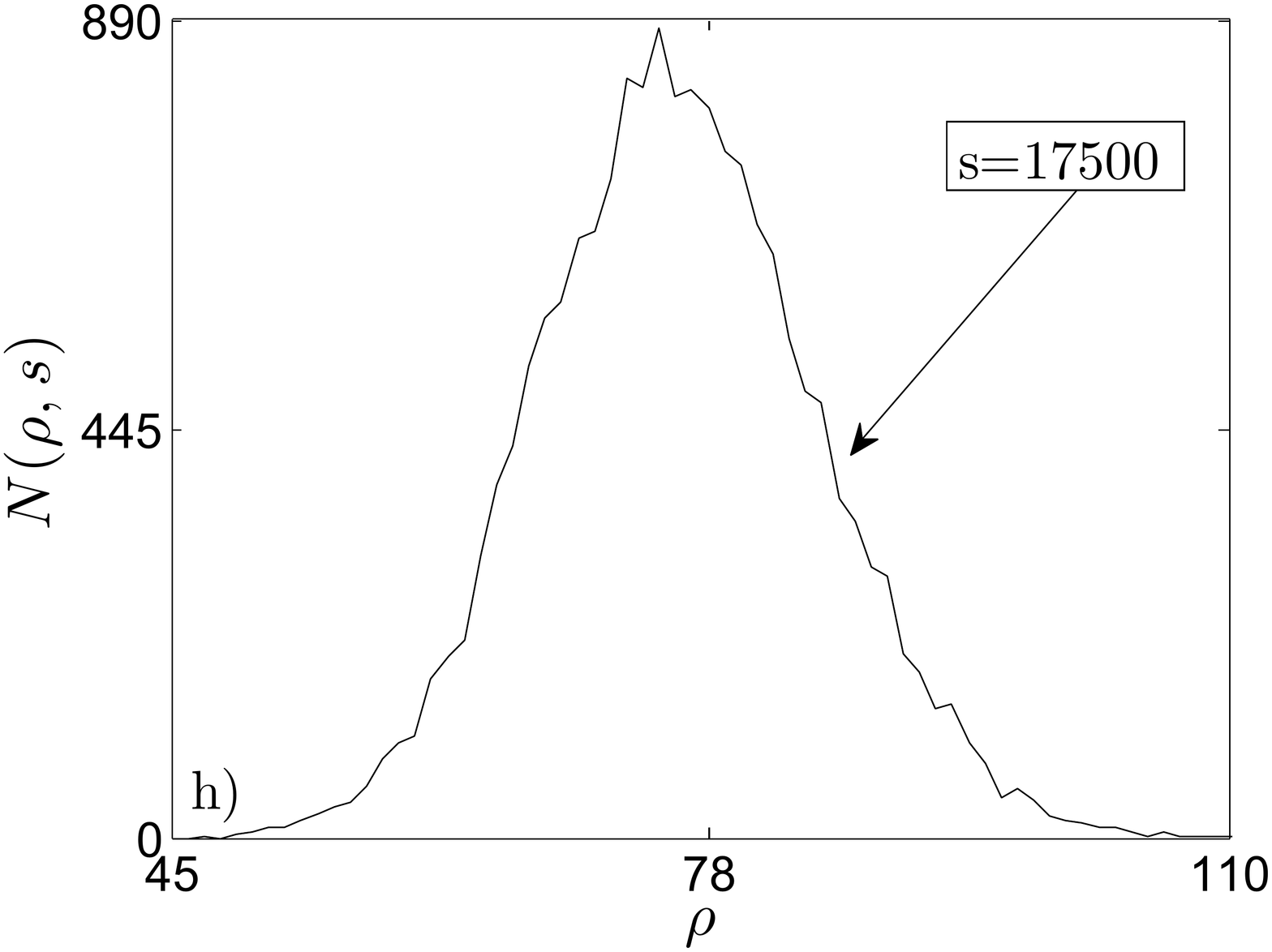}
\includegraphics[width=.21\textwidth, angle=270]{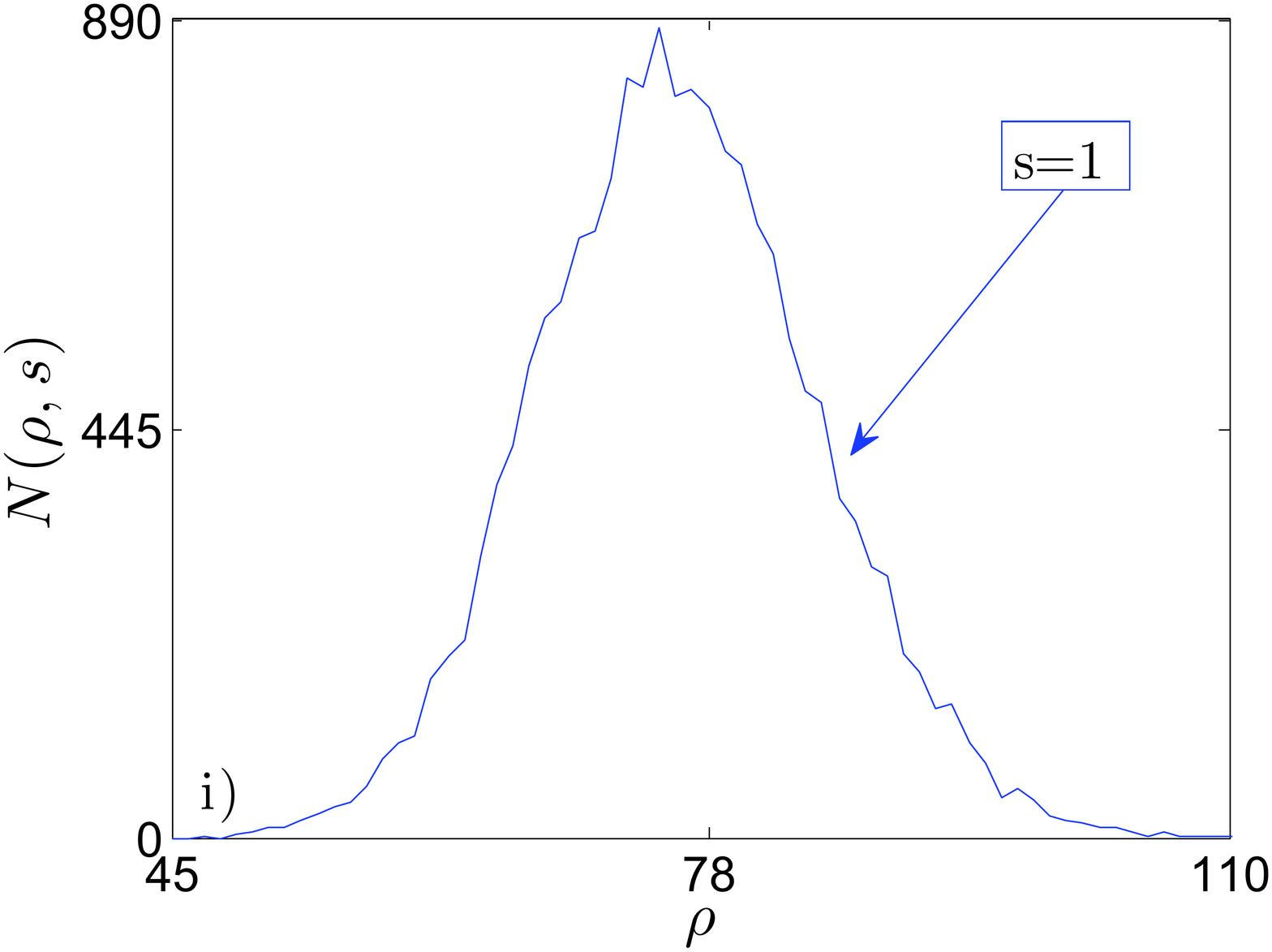}
\caption{\label{fig:Distr025}(Color on line) Number of nodes $N(\rho,s)$ corresponding to a string with $\rho$ non-null entries and belonging to a cluster of size $s$, plotted as a function of $\rho$, while different values of $s$ and shown in different colors. The system considered has size $V=17500$, $\theta=0.25$, $\alpha=0.1$ and $\gamma=1$. The arrangement of panels is the same as in the previous figure.}
\end{center}
\end{figure*}

Analogous results for the case $\theta=0.25$ are depicted in Fig.~\ref{fig:Distr025}. Now, in the intermediate regime, also for the WP a few small clusters emerge, while the abrupt jump in $S(f)$ evidenced for the SP is recovered here by the fact that $s$ assumes only two values: either $1$ or $V$.


We conclude this section with a remark. By focusing only on the topology of the graphs $\mathcal{G}(\alpha, \theta, \gamma, L)_{f,P}$, for a fixed parameter set and fixed $f$, we can compare the level of organization of the graph resulting from a different percolative process $P$. This can be attained by means of entropy measures \cite{ginestra}, which, given a particular ensemble, provide the normalized logarithm of the
number of networks in that ensemble, hence estimating how effective the features characterizing the ensemble are.
Here, we can fix the parameter set $(a, \theta, \gamma, L)$ and $f$ (this somehow fixes the ``energy" of the system) and measure the entropy within a configuration approach, namely working out the degree sequences; this approach is also related to a hidden variable model \cite{ginestra,AB}, consistently with the assignation of attributes.
For instance, for $\theta=0$, when the dilution is low (most links still present) the entropy of the ensemble WP is expected to be larger due to the presence of a few isolated nodes which yield a larger number of configurations; vice versa, when the dilution is increased the entropy of the ensemble RP is expected to prevail. Further analysis on this point may lead to speculate that a failure of a limited number of nodes is likely to involve only weak ties, while when the failure is wider it is more likely to involve any generic node. 
%
%
%
%
Similar reasoning can be extended in order to account also from an energy contribution due to the coupling.


\section{Clustering and correlations}\label{sec:correlation}
In this section we want to focus the attention on the properties of clustering and of correlation among links as we dilute them.

Before proceeding it is worth recalling that the clustering coefficient $C$ provides a measure of the transitivity of the graph and it can be calculated as the average over nodes $i$ of the local clustering coefficient $C_i$, defined as the actual number of links between the vertices within the neighborhood of $i$, divided by the maximum number of links that can exist between them, that is
\begin{equation} \label{eq:clustering}
C = \frac{1}{V} \sum_{i=1}^V C_i = \frac{1}{V} \sum_{i=1}^V \frac{2 E_i}{z_i (z_i -1)},
\end{equation}
where $E_i$ is the number of links among nodes which are connected to $i$ (node $i$ is not included), $z_i$ is the number of neighbors (also called degree) of $i$ and one conventionally sets $C_i=0$ for $z_i=0,1$.
A graph is often referred to as small-world, if its diameter is small (scaling as $\log V$, which is is verified by $\mathcal{G}$) and its average clustering coefficient  is significantly higher than the one relevant to a random graph constructed on the same vertex set, meaning $C=f$. As evidenced in \cite{AB,BA}, the graph under study can be defined as small-world.

\begin{figure}[tb] \begin{center}
\includegraphics[width=.45\textwidth]{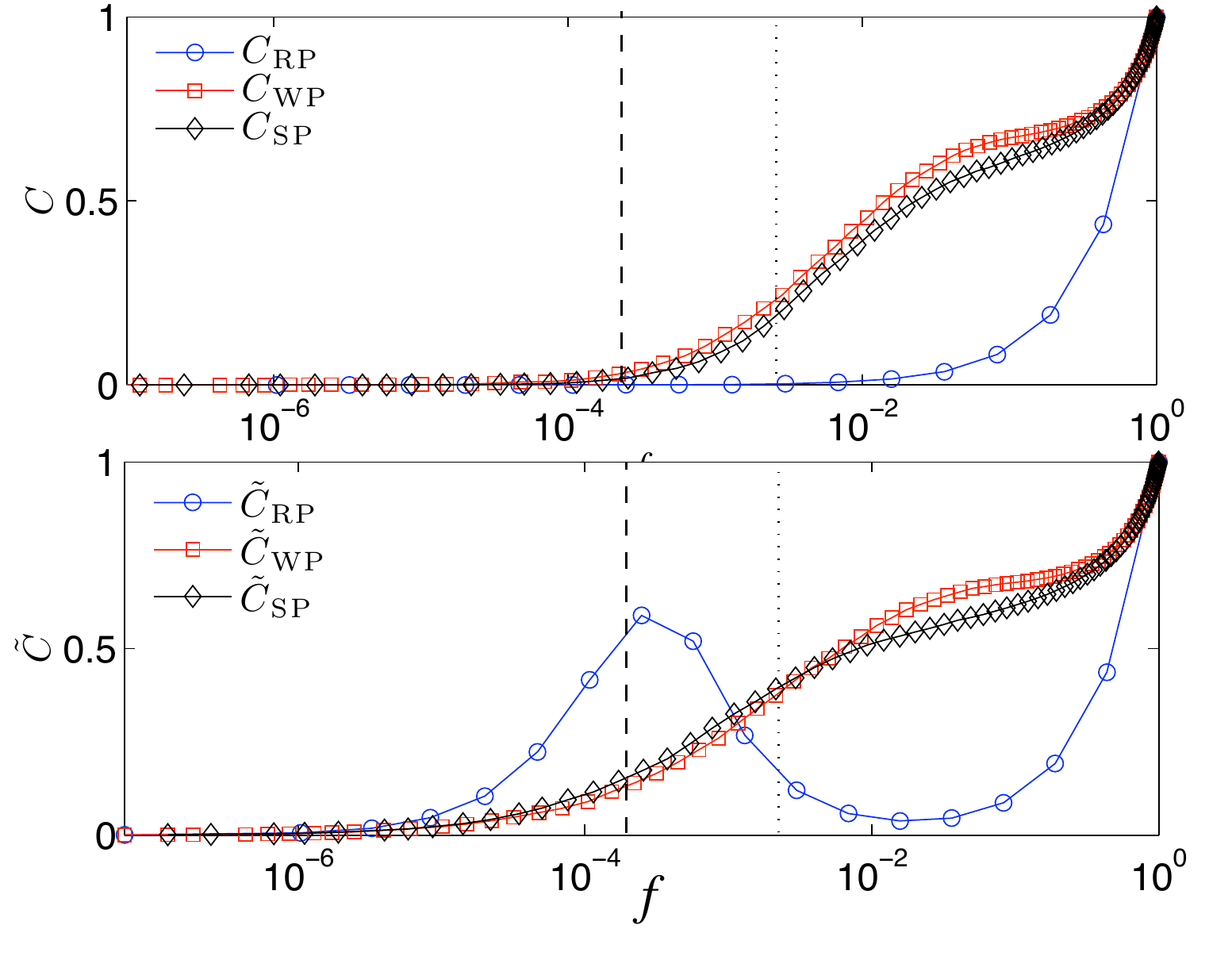}
\caption{\label{fig:C_0} (Color on line) Clustering coefficients $C$ (top panel) and $\tilde{C}$ (bottom panel) as a function of $f$ and for $\theta=0$, $V=5700$; different percolative processes are compared as explained by the legend. The vertical dashed and dotted lines highlight the percolation threshold for RP and WP, respectively.}
\end{center}
\end{figure}

As shown in Fig.~\ref{fig:C_0} (upper panel), when $\theta=0$, the clustering coefficient relevant to the graph obtained with a random dilution just corresponds, as expected, to the coefficient pertaining to an analogous ER graph; conversely, the clustering coefficients for deterministic dilution are larger and display a less trivial profile. 
In fact, for the RP, starting from small $f$ we first build up a set of uncorrelated small components having zero or very small coefficient, so that their contribution to $C$ is negligible; a significant and regular increase in $C$ is only set up from the percolation threshold (dashed line in the figures).

As for the WP, starting from small $f$ we first connect nodes having strings with large $\rho$ and these form a highly clustered component which already contributes to $C$; as $f$ is increased the largest component gradually expands and $C$ consistently grows; when $f \gtrsim f_c$ most nodes are connected and new links serve to connect low-degree nodes so that the rate of growth of $C$ is reduced; finally, when the network is connected (no isolated nodes) all new links determine an improvement in the clustering so that there is an acceleration in the growth of $C$.

Similar arguments apply for the SP percolation: for small $f$ only very weak ties are introduced and these, due to the relatively homogeneity of the graph ($a=0$), are sufficient to build up a structures component which progressively grows determining a larger and larger C.

\begin{figure}[tb] \begin{center}
\includegraphics[width=.45\textwidth]{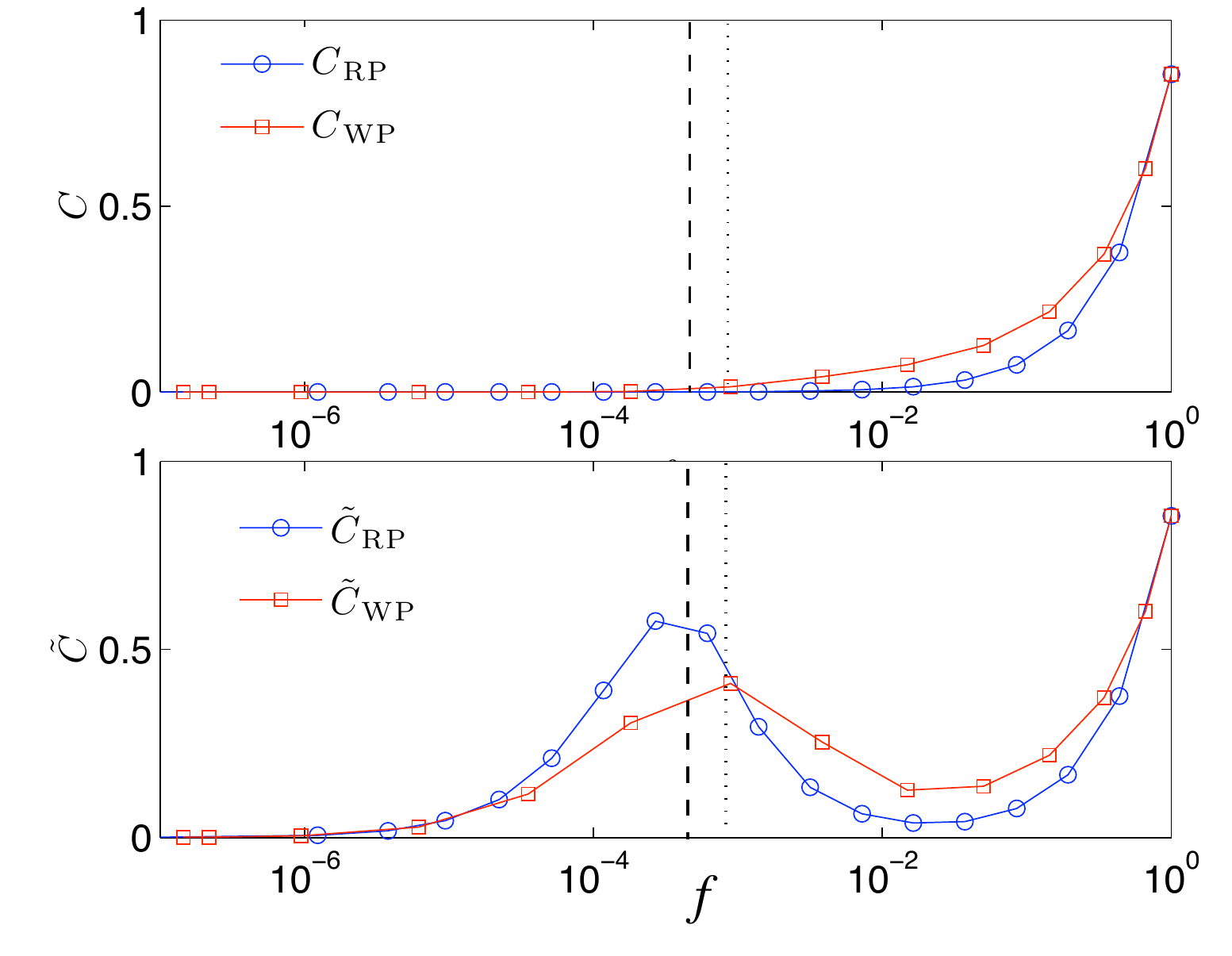}
\caption{\label{fig:C_025} (Color on line) Clustering coefficients $C$ (top panel) and $\tilde{C}$ (bottom panel) as a function of $f$ and for $\theta=0.25$, $V=5700$; different percolative processes are compared as explained by the legend. The vertical dashed and dotted lines highlight the percolation threshold for RP and WP, respectively.}
\end{center}
\end{figure}

In Fig.~\ref{fig:C_025} we show results for the case $\theta=0.25$; due to the finite-size effects affecting SP, we just focus on the cases RP and WP. Of course, for the RP no qualitative changes are evidenced with respect to the case $\theta=0$, while for the WP, as $f$ is increased, detached small clusters are now more likely to occur due to the sparsity of non-null entries in strings and this explains the fact that $C$ is now qualitatively comparable with the ER case.

Another interesting coefficient which we introduce, in order to monitor the evolving topology as links are removed, is a slightly modified version of the clustering coefficient, which we denote as $\tilde{C}$ and define as the average of the local dilution $\tilde{C}_i$, given by the fraction of links within the subgraph containing the node $i$ and all its neighbors, that is
\begin{equation}  \label{eq:dilution}
\tilde{C} = \frac{1}{V} \sum_{i=1}^V \tilde{C}_i =\frac{1}{V}  \sum_{i=1}^V \frac{2 \tilde{E}_i}{z_i (z_i +1)},
\end{equation}
where $\tilde{E}_i$ is the number of links connecting any couple of nodes belonging to the subgraph, including $i$ itself, and one conventionally sets $\tilde{C}=0$ for $z_i=0$. We remark that $\tilde{C}_i$ differs from $C_i$ by the fact that here we count all links within the neighborhood of $i$, including those stemming from $i$, i.e. $\tilde{E}_i = E_i + z_i$, from which $\tilde{C}_i=C_i+\sum_i 2/(z_i+1)$. Hence, once $f$ fixed, by comparing $C$ and $\tilde{C}$, one can derive information about the arrangement of existing links: either highly clusterized, so to form a small-sized connected component with nodes having relatively large degree (comparable $C$ and $\tilde{C}$) or highly scattered, so to eventually form a large connected component with nodes having relatively small degree (large $\tilde{C}$, small $C$).

As shown in Fig.~\ref{fig:C_0}, again for $\theta=0$ qualitative differences emerge between deterministic and random dilution. For deterministic percolations
$\tilde{C}$ follows a behavior similar to $C$: this results from the fact that, basically, we have only one component which keeps on growing as $f$ is increased, the remaining nodes being mainly isolated. Vice versa, for random percolation we can distinguish three different regimes, demarcated by two extremal points: in the first regime we have the emergence of several small components (e.g. dimers, trimers), each with small but non-null contributions to $\tilde{C}$; in the second regime such small components start to merge and this yields a reduction in $\tilde{C}$; finally, when the graph has reached a connected status, increasing the number of links can just produce an increase in $\tilde{C}$.  

For $\theta=0.25$, also the WP transition display the same multi-regime behavior, which confirms the picture above.
%

\section{Conclusions and Perspectives}\label{sec:conclusions}
In this work we analyzed the evolution, under percolation process, of a class of weighted graphs $\mathcal{G}$ introduced in \cite{AB,BA}, whose topological properties arise from imitative interaction among nodes and can be properly varied by tuning the parameters ($\alpha$, $\theta$, $\gamma$, $L$).
In particular, here we fixed $\alpha=0.1$, $\gamma=1$ and $\theta = 0$ or $\theta = 0.25$, while the size ($V= L/ \alpha$ is the number of nodes) is varied; such a situation corresponds, for large enough volumes, to fully-connected graphs, still retaining a non-trivial distribution $P_{\mathrm{coupl}}(J; \alpha, \theta,\gamma,L)$ for the coupling strength $J$ associated to any link. This allows to perform and compare different percolation processes: random (RP, where links are randomly extracted for deletion), deterministic-weak (WP, where links are deleted starting from the weakest ones) and deterministic-strong (SP, where links are deleted starting from the strongest ones).

Our results highlight that weak ties are the most crucial in order to ensure the overall connection of the system, 
that is, the size of the largest component starts to be smaller than $V$ when only few (weak) links are deleted. 
When $\theta$ approaches value $0.5$ from below (and in the presence of finite-size effects which affect the skewness of the coupling distribution) one can see that the spanning tree underlying $\mathcal{G}$ is mostly made up of weak links, while strong linkes are unlikely and mainly redundant. 
Hence the robustness of $\mathcal{G}$ sensitively depends on which ties are the most prone to failure. The fact that weak ties are fundamental to maintain the whole graph connected is consistent with the so-called theory of weak-ties \cite{granovetter,noi}, according to which, in social systems, weak ties work as bridges between different sub-communities.

Moreover, we showed that removing in rank order, from the weakest to the strongest ties, shrinks the network, but does not precipitously break it apart, in such a way that the percolation is rather smooth. 
A similar phenomenon has been evidenced in the context of social networks where, when all ``declared friendships'' are considered the graph is highly connected, but when only ``strong'' links are retained, selecting firstly ``maintained'' relationships and secondly ``mutual'' relationships, nodes get gradually disconnected forming only small subclusters \cite{fb}. Conversely, as we underlined, RP gives rise to more structures subclusters while diluting.

%
A possible extension of this work could consider non-complete graphs ($\theta>0.5$) with random deletion of nodes so to evaluate whether also for such correlated networks, degree-degree correlation yields qualitative changes in the percolation behavior as expected from \cite{referee}.

Analysis similar to those performed here can involve different connecting rules (see Eq.~\ref{eq:rule}, \cite{immuno}) in order to figure out a possible relation between the kind of interaction (e.g. imitative or anti-imitative) and the dynamic behavior. Also, a possible mapping between the dilution obtained via cutting a fraction $1-f$ of links and via a progressive reduction of the parameter $a$ may be figured out.

\bigskip

\section*{Acknowledgments}
The authors are grateful to Adriano Barra and to Daniele Del Sarto for interesting discussions and suggestions.\\
This work is supported by FIRB grant $RBFR08EKEV$.

\appendix \section{Degree-degree correlation} \label{appe1}
In this appendix we aim to show that overpercolated networks $\mathcal{G}$ display negative assortativity by calculating how the average degree of nodes belonging to the neighborhood of $i$ depends on the degree of $i$ itself.
In fact, we can write that in a graph $\mathcal{G}(\alpha,\theta, \gamma,L)$ the probability for a node $i$ to have neighbors which display in the average $z$ neighbors is 
\begin{eqnarray}
\nonumber
&&P_{\textrm{deg-deg}}(z;\rho_i,a,L) = \\
\nonumber
&&\frac{1}{\mathcal{N}}\sum_{\rho_j=1}^L P_{\textrm{link}} (\rho_i,\rho_j,L) P_{\textrm{deg}} (z; \rho_j,a,V) P_1(\rho_j;a,L),
\end{eqnarray}
where $\mathcal{N}$ is the normalization factor, $P_{\textrm{link}} (\rho_i,\rho_j,L) = 1 - P_{\textrm{match}} (0; \rho_i,\rho_j,L)$ is the probability that there exists a link connecting $i$ and $j$ and $P_{\textrm{deg}} (z; \rho_j,a,V)= \binom{V}{z} [\bar{P}_{\textrm{link}}(\rho_j;a)]^{z}[1-\bar{P}_{\textrm{link}}(\rho_j;a)]^{V-z}$ is the probability that node $j$ has $z$ neighbors.
Hence one finds that the average degree for $i$'s neighbors is
\begin{eqnarray}
\nonumber
&&\tilde{z}(\rho_i;a,L) = \sum_{z=0}^V P_{\textrm{deg-deg}}(z;\rho_i,a,L)  z = \\
\nonumber
&&\frac{1}{\mathcal{N}}\sum_{\rho_j=1}^L P_{\textrm{link}} (\rho_i,\rho_j,L) P_1(\rho_j;a,L) \bar{z}(\rho_j;a,L,V),
\end{eqnarray}
being $\bar{z}(\rho_j;a,V) = V \{1- [(1-a)/2]^{\rho_j} \}$ the average degree for node $j$.
With some algebra one gets to
\begin{eqnarray}
\nonumber
&&\tilde{z}(\rho_i;a,L) = 1- \left ( \frac{1-a}{2} \right) ^L \left( \frac{3+a}{2} \right)^L  \\
&\times& \left[  1 - \left( \frac{2}{3+a} \right)^{\rho_i} \right] \left[  1 - \left( \frac{1-a}{2} \right)^{\rho_i} \right]^{-1}.
\end{eqnarray}
Now, noticing that $0<(1-a)/2 < 2/(3+a)<1$, we can deduce that $\tilde{z}(\rho_i;a,L,V)$ is decreasing with $\rho_i$, namely with $\bar{z}(\rho_i;a,L,V)$, so that, as long as the mean-field approach developed here is valid \cite{AB,BA}, the graph displays dissortativity.

\section{Analytical Results on RP} \label{appe2}

\begin{figure}[tb]
\includegraphics[height=60mm]{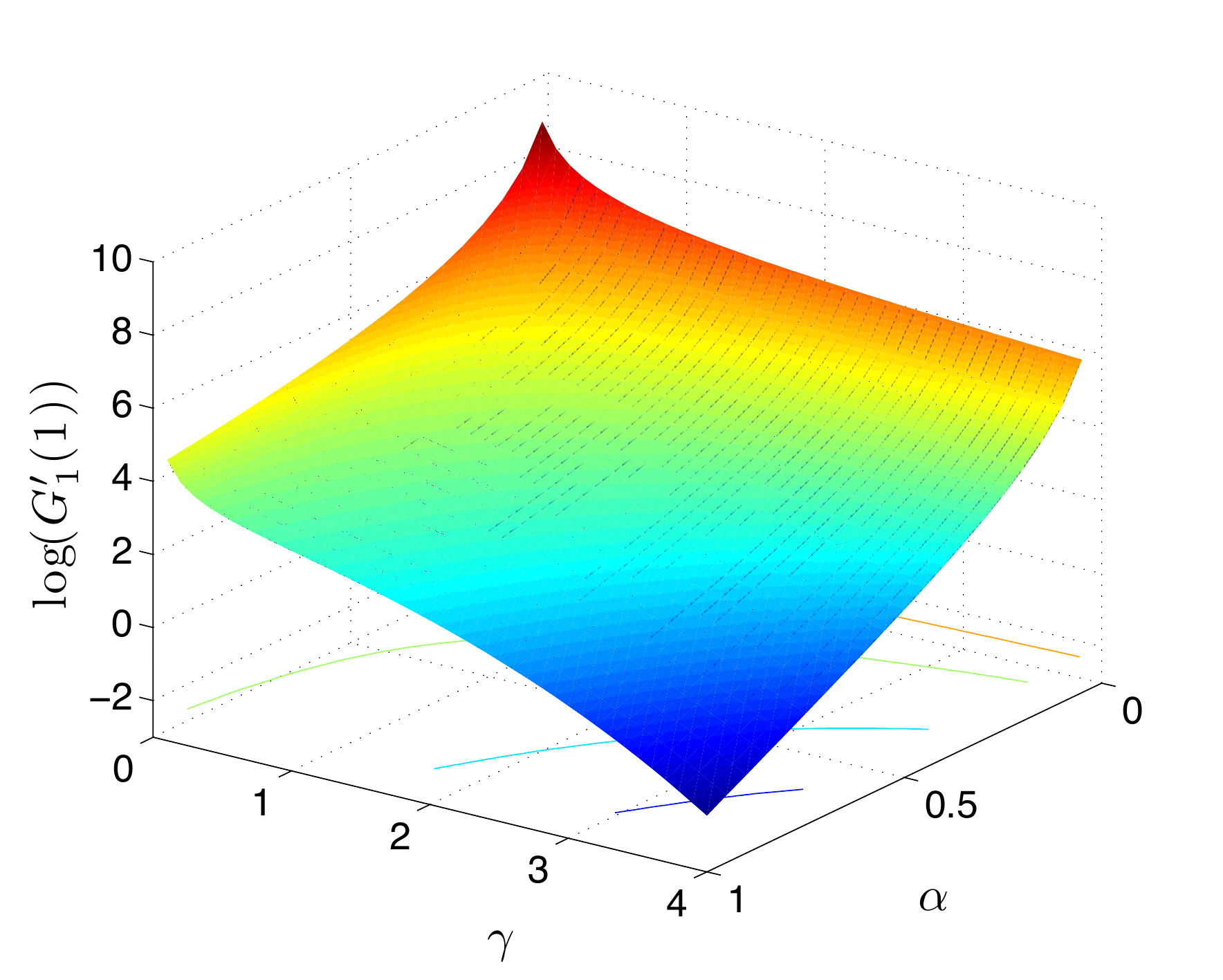}
\caption{\label{fig:generating} \label{fig:weak}
(Color on line) Natural logarithm of $G_1'(1)$ for $L=20$ , $\theta=0.5$ and different values of
$\gamma$ and $\alpha$ as given by Eq.~\ref{eq:G1}. 
Whenever $G'(1) >1$ a giant component emerges.}
\end{figure}

The percolation problem has been studied over different kinds of
structure, both analytically and numerically
\cite{percola1,percola2}; in particular, within the so-called
configuration model approach \cite{newman,albert}, we can exploit
the generating function formalism to get some insights into the
problem. First of all, being $\bar{P}_{\mathrm{degree}}(k)$ the average degree distribution for the generic graph $\mathcal{G}$ (here we drop the dependence on the parameter set to lighten the notation), we define
\begin{equation}
G_0(x) = \sum_{k=0}^{\infty} \bar{P}_{\mathrm{degree}}(k) x^k, \; \: G_1(x) = \sum_{k=0}^{\infty} Q(k) x^k,
\end{equation}
where $Q(k) = (k+1) \bar{P}_{\mathrm{degree}}(k+1) / \bar{z}$. Now, assuming $V$ large and the clustering not significantly, 
$Q(k)$ is the so-called excess degree \cite{newman}, representing the degree
distribution of the vertex at the end of a randomly chosen edge;
notice that $G_0'(1)=\bar{z}$ and $G_1(x) = G_0'(x)/\bar{z}$.
Recalling that $\bar{P}_{\mathrm{degree}}(k) = \sum_{\rho=0}^{L}
\binom{V}{k} \left[ 1 - \left(
\frac{1-a}{2}  \right)^{\rho} \right]^k   \left( \frac{1-a}{2}
\right)^{\rho(V-k)}  P_1(\rho;a,L)$, (see \cite{AB,BA}), we can write
\begin{equation}
\nonumber
G_0(x) = \sum_{\rho=0}^L P_1(\rho; a, L) \left[ x + (1-x) \left( \frac{1-a}{2} \right)^{\rho} \right]^V.
\end{equation}
Moreover,  for uniform link deletion probability, the mean cluster
is \cite{callaway}
\begin{equation}
\bar{s} = 1 + f G_0'(1) + \frac{f^2 G_0'(1)}{1 - f G_1'(1)},
\end{equation}
which diverges when $1 - f \, G_1'(1)=0$; this point marks the percolation threshold of the system: for $f > f_c = 1/G_1'(1)$ a giant component of connected vertices is established.
Therefore, consistently with the Molloy-Reed criterion \cite{molloy}, when $G_1'(1)<1$ the graph consists of many small components, while when $G_1'(1)>1$ a giant component can emerge.
Here we find
\begin{equation}
G_1'(1) =  \frac{G_0''(1)}{\bar{z}} = \frac{\bar{z}}{[1 - h(a)]^2} \left[ 1 - 2 h(a) + g(a)\right],
\end{equation}
where $h(a) = [(3 - 2a - a^2)/4]^L = p$ and $g(a)=[(1-a)(5-a^2)/8]^L$. Assuming $a = -1 +\gamma (\alpha/L)^{\theta}$ and posing $\tilde{\gamma} = \gamma (\alpha/L)^{\theta} / 2$, we can write
\begin{eqnarray}\label{eq:G1}
\nonumber
G_1'(1) &=& \frac{\bar{z}}{\left[1 - (1 - \tilde{\gamma}^2)^L \right]^2} \Big[1 - 2 (1- \tilde{\gamma}^2)^L  \\
&+& (1 - 2 \tilde{\gamma}^2 + \tilde{\gamma}^3)^L \Big]  \underset{L \to \infty}{\to} \bar{z},
\end{eqnarray}
in analogy with the percolation threshold expected for the ER graph.
In Fig.~\ref{fig:generating} we show $G_1'(1)$ as a function of $\gamma$ and $a$ and for a finite value of $L$.

Finally, the formalism developed  in the first part of this
appendix has also been used to find an expression for the global
clustering coefficient or transitivity of the configuration model \cite{pnas}:
\begin{equation} \label{eq:clustering}
c \equiv \frac{3 N_{\triangle}}{N_3}= \frac{\bar{z}}{V} \left[ \frac{ \bar{z^2} - \bar{z} }{\bar{z}^2}  \right]^2 = \frac{1}{V z} \left( \frac{G_0'(1) G_1'(1)}{z}  \right)^2,
\end{equation}
where $N_{\triangle}$ is the number of triangles in the network
and $N_3$ is the number of connected triples of vertices
\cite{newman}. Notice that from Eq.~\ref{eq:clustering}, $c$ is
given by the coefficient expected for the ER graph, namely
$\bar{z}/V$, times an extra factor such that when the degree
distribution is highly skewed, given that the factor $\bar{z^2}/
\bar{z}^2$ can be rather large, $c$ is not necessarily negligible
for the graph sizes relatively large. Interestingly, we find
\begin{equation}
c = \frac{[1 - 2h(a) + g(a)^2]^2}{[1-h(a)]^3},
\end{equation}
which, for $a \in [-1, 1]$ is always larger than $p=1-h(a)$, hence confirming the large degree of cliquishness of $\mathcal{G}$.


\begin{thebibliography}{}
%
\bibitem{reviews}
R. Albert, A.-L. Barab\'asi, \emph{Rev.\ Mod.\ Phys.}, \textbf{74}, 47 (2002);
S.N. Dorogovtesev, J.F.F. Mendes, \emph{Adv. Phys.}, \textbf{51}, 1079 (2002);
M.E.J. Newman, \emph{SIAM Rev.}, \textbf{45}, 167 (2003)
%
%
\bibitem{libro}
H.D. Rozenfeld, \emph{Structure and Properties of Complex Networks: Models, Dynamics, Applications} (VDM Verlag, 2008)
%
\bibitem{kiss} H.J.M. Kiss, A.M. Mihalik, T. N\'an\'asi, B. \"Ory, Z. Spir\'o, C. S\"oti and P. Csermely, BioEssay \textbf{31}, 651 (2009)
%
\bibitem{ACV} E. Agliari, M. Casartelli and A. Vezzani, J.\ Stat.\ Mech., P10021 (2010).
%
\bibitem{stauff} D. Stauffer and A. Aharony, \emph{Introduction to percolation theory} (Taylor \& Francis, London 1994)
%
\bibitem{essam} J.W. Essam, Rep.\ Prog.\ Phys.\ \textbf{43}, 53 (1980)
%
\bibitem{palla} G. Palla, I. Derenyi, T. Vicsek, Phys.\ Rev.\ E \textbf{69}, 046117 (2004).
%
\bibitem{appl1} I. Breskin, J. Soriano, E. Moses, T. Tlusty, Phys.\ Rev.\ Lett. \textbf{97}, 188102 (2006)
\bibitem{appl2} L. Huang, Y.-C. Lai, K. Park, J. Zhang, Phys.\ Rev.\ E \textbf{73}, 066131 (2006)
\bibitem{appl3} Y. Chen, G. Paul, R. Cohen, S. Havlin, S.P. Borgatti, F. Liljeros, H.E. Stanley, Physica A \textbf{378}, 11 (2007)
\bibitem{appl4} M. Brede, U. Behn, Phys.\ Rev.\ E \textbf{67}, 031920 (2003)
%
\bibitem{BA}
A. Barra, E. Agliari, \emph{Equilibrium statistical mechanics on correlated random graphs}, J.\ Stat.\ Mech.\ P02027 (2011)
\bibitem{AB}
E. Agliari, A. Barra, \emph{A Hebbian approach to complex network generation}, Europhys. Lett. 94 10002 (2011)
%
\bibitem{bio1} E. Bullmore, O. Sporns, Nature \textbf{10}, 186 (2009)
\bibitem{bio2} T. Yamada, P. Bork, Nature \textbf{10}, 791 (2009)
\bibitem{techno} C. Zhao, Intelligent Computing and Information Science \textbf{135}, 124 (2011)
\bibitem{socio} L.M. Sander, C.P. Warren, I.M. Sokolov, Physica A \textbf{325}, 1 (2003)
%
\bibitem{dissorta} H.-B. Hu, X.-F. Wang, Europhys. Lett. \textbf{86}, 18003 (2009)
\bibitem{fb} D. Easley, J. Kleinberg, \textit{Networks, Crowds, and Markets}, Cambridge University Press (2010)


\bibitem{amit} D.J. Amit, \emph{Modeling brain functions. The world of attractor neural networks}, (Cambridge Press, 1988).
%
\bibitem{doro} R.A. da Costa, S.N. Dorogovstev, A.V. Goltsev, J.F.F. Mendes, archive:1009.2534.
%
\bibitem{ER}
P. Erd\"{o}s, A. R\'enyi, Publicationes Mathematicae \textbf{6}, 290�297 (1959).
%
\bibitem{nature}
R. Albert, H. Jeong, A.-L.  Barab\'{a}si, \emph{Error and attack tolerance of complex networks}, Nature \textbf{79} 378 (2000)
%
\bibitem{immuno} A. Barra and E. Agliari, \emph{Statistical mechanics approach to autopoietic immune networks}, J.\ Stat.\ Mech.\ P07004 (2010).
%
\bibitem{onnela} J.-P. Onnela, J. Saram\"aki, J. Hyv\"onen, G. Szab\'o, D. Lazer, K. Kaski, J. Kert\'esz and A.-L. Barab\'asi, PNAS \textbf{104}, 7332 (2007)

%

\bibitem{bela}
B. Bollob\'as, \emph{Random graphs}, Cambridge University Press, Cambridge (2001)



\bibitem{ginestra} G. Bianconi, Phys.\ Rev.\ E \textbf{79}, 036114 (2009).
\bibitem{granovetter} 
M.S.  Granovetter, {\em The Strength of Weak Ties}, Amer. J. of Sociology \textbf{78},
$1360-80$, (1973).
%
\bibitem{noi} A. Barra, E. Agliari, \emph{A statistical mechanics approach to Granovetter theory}, to appear on Physica A.

\bibitem{referee}
M.E.J. Newman, \emph{Assortative Mixing in Networks}, Phys.\ Rev.\ Lett., \textbf{89}, 208701 (2002)
%

%
\bibitem{percola1}
G.R. Grimmett, \emph{Percolation}, Spinger-Verlag, Berlin 1999.
\bibitem{percola2}
B. Bollob\'{a}s, O. Riordan \emph{Percolation}, Cambridge University Press, Cambridge 2006.
%
\bibitem{albert}
R. Albert and A.-L. Barab\'{a}si, \emph{Statistical mechanics of Complex Networks}, Rev. Mod. Phys. 74, 47-98 (2002), and references therein
%
\bibitem{newman}
M.E.J. Newman, \emph{The structure and function of complex networks}, SIAM Review, \textbf{45}, 167-256 (2003), and references therein
%
\bibitem{callaway} D. Callaway, M.E.J. Newman, S.H. Strogatz, D.J. Watts, {\em Network robustness and fragility: Percolation on random graphs}, Phys. Rev. Lett. \textbf{85}, 5468 (2000).
%
\bibitem{molloy}
M. Molloy and B. Reed, \emph{The size of the giant component of a random graph with a given degree sequence}, Combinatorics, Probability and Computing 7, 295-305 (1998)









\bibitem{pnas}
M.E.J. Newman, D.J. Watts, S.H. Strogatz, Proc.\ Nat.\ Am.\ Soc., \textbf{99}, 2566 (2002)



%
%
%
%
%
%






\end{thebibliography}
\end{document}